\documentclass[structabstract]{aa}  
\usepackage{graphicx}
\usepackage{natbib}
\usepackage[table]{xcolor}
\usepackage{amssymb}
\usepackage{lscape}
\bibpunct{(}{)}{;}{a}{}{,} % to follow the A&A style

\begin{document}

   \title{The star cluster - field star connection in nearby spiral galaxies.}
   \subtitle{II. Field star and cluster formation histories and their relation.\thanks{Table 10 is only available in electronic form
at the CDS via anonymous ftp to cdsarc.u-strasbg.fr (130.79.128.5)
or via http://cdsweb.u-strasbg.fr/cgi-bin/qcat?J/A+A/}}
   \author{E. Silva-Villa
          \and
          S. S. Larsen
          }

   \institute{Astronomical Institute, University of Utrecht,
              Princetonplein 5, 3584 CC, Utrecht, The Netherlands\\
              \email{[e.silvavilla,s.s.larsen]@uu.nl}
         			}
 							
%  \date{Received September 15, 1996; accepted March 16, 1997}

  \abstract 
% context heading (optional) % {} leave it empty if necessary
	{Recent studies have started to cast doubt on the assumption that most stars are formed in clusters. Observational studies of field stars and star 
	cluster systems in nearby galaxies can lead to better constraints on the fraction of stars forming
	in clusters. Ultimately this may lead to a better understanding of star formation in galaxies, and
	galaxy evolution in general.}
% aims heading (mandatory) 
	{We aim to constrain the amount of star formation happening in long-lived clusters for four galaxies through the homogeneous, simultaneous study of field stars and star clusters.} 
% methods heading (mandatory) 
	{Using HST/ACS and HST/WFPC2 images of the galaxies NGC~45, NGC~1313, NGC~5236, and NGC~7793,
	we estimate star formation histories by means of the synthetic CMD method. Masses and ages of star clusters are estimated using simple stellar population
	model fitting. Comparing observed and modeled luminosity functions, we estimate cluster formation rates.
	By randomly sampling the stellar initial mass function (SIMF), we construct artificial star clusters and 
	quantify how stochastic effects influence cluster detection, integrated colors, and age estimates.}
% results heading (mandatory) 
	{Star formation rates appear to be constant over the past $10^7-10^8$ years within the fields covered 
	by our observations. The number of clusters identified per galaxy
	varies, with a few detected massive clusters (M $\ge 10^5$ M$_{\odot}$) and a few older than 1 Gyr. Among our sample of galaxies,
	NGC~5236 and NGC~1313 show high star and cluster formation rates, while NGC~7793 and NGC~45 show lower values.
	We find that stochastic sampling of the SIMF has a strong impact on the estimation of ages, colors, and completeness for clusters with masses $\le 10^3-10^4$ M$_{\odot}$, 
	while the effect is less pronounced for high masses. Stochasticity also makes size measurements highly uncertain at young ages ($\tau\lesssim 10^8$ yr), making it difficult to distinguish between 
	clusters and stars based on sizes.} 
% conclusions heading (optional), leave it empty if necessary 
	{The ratio of star formation happening in clusters ($\Gamma$) compared to the global star formation 
	 appears to vary for different galaxies. We find similar values to previous studies
	 ($\Gamma \approx$ 2\%--10\%), but we find no obvious relation between
	 $\Gamma$ and the star formation rate density ($\Sigma_{\rm SFR}$) within the range  
	 probed here ($\Sigma_{\rm SFR} \sim 10^{-3} - 10^{-2} M_\odot$ yr$^{-1}$ kpc$^{-2}$). 
	 The $\Gamma$ values do, however, appear to correlate with the specific U-band luminosity (T$_L$(U), the fraction of total light coming from clusters compared to the 
	 total U-band light of the galaxy).}
   \keywords{
   	   galaxies: Individual -- NGC~5236
   	   galaxies: Individual -- NGC~7793
   	   galaxies: Individual -- NGC~1313
   	   galaxies: Individual -- NGC~45
   	   galaxies: Star clusters --
             galaxies: Star formation --
             galaxies: Photometry
             }
   \maketitle
%
%________________________________________________________________

%%%%%%%%%%%%%%%%%%%%%%%%%
%INTRODUCTION
%%%%%%%%%%%%%%%%%%%%%%%%%

\section{Introduction}
It is often assumed that stars can be formed either in the field of a galaxy as single stars or in a group of stars (cluster) that formed from the same
molecular cloud at the same time. However this view has been recently questioned by \citet{bressert10}, who challenged the idea that field and 
cluster formation are actually distinct modes of star formation.

Owing to dynamical and stellar evolution clusters disrupt \citep{spitzer87} and the stars become members of
the field stellar population. 
While it is commonly assumed that most (if not all) stars formed in 
clusters \citep[e.g.][in the solar neighborhood]{LL03,porras03}, the amount 
of star formation happening in those clusters that remain bound beyond the embedded phase is still uncertain. 
In any case, if clusters are to be used as tracers of
galactic star formation histories, it is of key importance to understand what fraction of star 
formation is happening in long-lived clusters and whether or not this fraction correlates
with other host galaxy parameters. Following \citet{bastian08}, we refer to this
fraction as $\Gamma$.

Estimating $\Gamma$ is not straightforward. Apart from differences arising at the time of
formation, cluster disruption will also affect the detected number of clusters of a
given age ($\tau$). 
\citet{LL03} estimate that between 70\% to 90\% of the stars in the solar neighborhood
form in embedded star clusters, while only 4--7\% of these clusters survive for more than
about 100 Myr. Similarly, \citet{lamersgieles08} estimate an ``infant mortality'' rate
of 50\% to 95\%, based on a comparison of the surface density of open clusters and the
star formation rate near the Sun. By studying the UV flux in and out of clusters in the galaxy 
NGC 1313, \citet{pellerin07} suggest that over 75\% (between 75\% and 90\%) of the flux is 
produced by stars in the field, concluding that the large number of B-type stars in the field 
of the galaxy could be a consequence of the (high) infant mortality of clusters.
For the Small Magellanic Cloud, \citet{gielesbastian08} estimate that optically
visible, bound clusters account for 2\%-4\% of the star formation, while \citet{gieles09} estimates 
this fraction to be in the range 5\%--18\% for the spiral galaxies M74, M51, and M101. However,
most of these studies could not distinguish between scenarios in which a large fraction of stars initially
form in clusters that rapidly dissolve or whether there is a genuine ``field'' mode of
star formation. Studying the Antennae galaxy, \citet{fall04} estimates that 20\% (and possibly all) 
stars were formed in clusters. Using a larger sample of galaxies, \citet{goddard10} find a power-law 
relation ($\Gamma\propto\Sigma_{\rm SFR}^{\alpha}$) between the fraction of stars forming in clusters 
that survive long enough to be optically visible
and the star formation rate density of the galaxy 
($\Sigma_{\rm SFR}$).
Their data set covers different types of galaxies, from irregulars (i.e. LMC, SMC, and NGC~1569)  
to grand design spirals (i.e. NGC~5236).
The $\Sigma_{\rm SFR}$ of these galaxies vary from $7\times10^{-3}$ to $\sim700\times10^{-3}$ M$_{\odot}$yr$^{-1}$Kpc$^{-2}$.
Their $\Gamma$ vs.\ $\Sigma_{\rm SFR}$ relation, however, is based on somewhat heterogenous
data with different mass- and age 
ranges, which do not come from the same observations \citep[see details in Sect. 4 of][]{goddard10},
although the authors do attempt to homogenize the sample by normalizing the
cluster samples to a common mass limit.

The actual definition of the phase called {\em infant mortality} is somewhat ambiguous in the 
literature. Early disruption due to rapid gas expulsion may only take a few Myrs, but the term
has also been used to describe mass-independent disruption, meaning that clusters lifetime is independent of mass over a much
longer time span. In the latter case, the ``infant mortality rate'' (IMR) refers to the fraction
of clusters that are disrupted per decade of age. We prefer to simply use the term 
``mass-independent'' disruption (MID) in this case. For MID, the IMR is related to the slope $a$ of the 
age distribution, $dN/d\tau \propto \tau^{a}$ of a mass limited cluster sample 
as $a = \log(1-{\rm IMR})$ \citep{whitmorechandarfall07}. 
For example, \citet{degrijs08} found that for the SMC the IMR is close to 30\% (between 3-160 Myr),
while the logarithmic age distribution of clusters in the Antennae galaxies is about flat
($a\approx-1$),
indicating an IMR close to 90\% \citep{fall04}, assuming that the star formation rate has been
about constant over the past few $10^8$ years. 
On theoretical grounds, the time scale for the (gradual) cluster disruption is 
expected to be mass-dependent, owing to tidal shocks and evaporation that follows early gas expulsion
\citep[e.g.][]{gieles06}, assuming there is no strong relation between cluster mass and radius. 
In this description, the dissolution time $t_{\rm dis}$ of a cluster scales with cluster mass
as $t_{\rm dis} = t_4 (M/10^4 M_\odot)^\gamma$, where $t_4$ is the lifetime of a $10^4$ M$_\odot$
cluster \citep[see][]{BL03,lamers05}. The time scale on which clusters
dissolve may also depend on external factors, such as the tidal field strength, density of molecular gas, passages near/through giant molecular clouds, or through spiral arms, etc.
\citep[see e.g.][]{gieles06,gieleslamers07}. This scenario attempts to compile in one single formula
all the possible processes that affect cluster disruption.
See \citet{lamers09} for a description of the different models for cluster
dissolution.
 
Determining the extent to which cluster dissolution is a mass-dependent process has turned out to be 
difficult. Estimations of cluster parameters based on observations are affected by stochastic effects, 
degeneracies, and observational uncertainties. For example, \citet{maizapellaniz09} used Monte 
Carlo simulations to estimate how stochastic effects coming from the random sampling of the 
stellar initial mass function influence the determination of ages and masses, which are derived 
from broadband photometry. 
\citet{piskunov09} show how the consideration of the discreteness of 
the stellar initial mass function (IMF) can explain features observed in the color-age relation and can 
improve the fit between models and observations. They conclude that the large number of red outliers
can be explained as a systematic offset coming from the difference between discrete- and continuous-IMF
at low masses (M$_c$=$10^2$ M$_{\odot}$) and young ages (log$(\tau)[yr]\sim$7), reaching up to $\sim$0.5 magnitudes, and decreases down 
to $\sim$0.04 magnitudes at higher masses (M$_c$=$10^6$ M$_{\odot}$).

\begin{figure}[!t]
	\centering
		\includegraphics[trim= 0mm 0mm 0mm 0mm,width=\columnwidth]{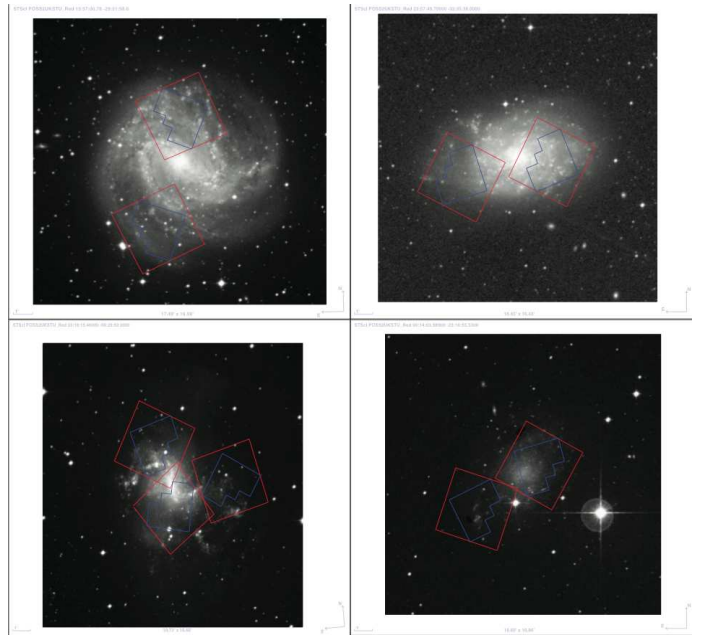}
		\caption{Galaxies studied in this paper. {\it Top left}: NGC~5236; {\it top right}: NGC~7793; {\it bottom left}: NGC~1313; and {\it bottom right}: NGC~45.
			      Red lines represent the pointings covered by the HST/ACS, while the blue lines represent the 
			      pointings of the HST/WFPC2. Images were taken
			      from the DSS archive using Aladin software.}
	\label{fig:set}
\end{figure}

To estimate field star formation histories, a different approach is needed than for
clusters, because ages cannot
in general be determined directly for individual stars. \citet{tosi91} presented 
a method that takes incompleteness, resolution, depth, and observational errors 
(among other parameters) into account to construct a synthetic color-magnitude diagram (CMD),
which can be used to estimate the star formation history by comparison with observations . This method has 
been developed further by other authors in the past years, 
e.g. \citet{dolphin97} and \citet{harriszaritsky01}, and has been used for a large number of galaxies, 
e.g. SMC, LMC \citep{harriszaritsky04,harriszaritsky09}, M31 \citep{brown08}, NGC~1313
\citep{larsen07}. In this series of papers, we make use of this method
to estimate the field star formation rates of our target galaxies, which we then compare
with cluster formation rates to estimate $\Gamma$.

In \citet[][hereafter Paper I]{silvavilla10}, we presented the tools needed to study and constrain 
the $\Gamma$ value of our set of galaxies, and used NGC~4395 as a testbed galaxy.
As the second paper in a series, this paper aims to estimate $\Gamma$ in different environments and 
compare it with previous work \citep[e.g.][]{gieles09,goddard10}, using the
complete set of galaxies. To this end, we took advantage of the superb spatial resolution of the 
{\em Hubble Space Telescope} (HST) and used images of the galaxies NGC~5236, NGC~7793, NGC~1313, and NGC~45, 
which are nearby, face-on spiral galaxies that differ in their current star formation rates and morphology. 
These galaxies are near enough ($\approx 4$ Mpc)
to allow us to disentangle the cluster system from the field stars, making it possible to estimate 
cluster and star formation histories separately and simultaneously from the same data.

\begin{table*}[!t]
	\centering
	\caption{Galaxy parameters.}
		\begin{tabular} {c c c c c c} 
			\hline \hline
			Galaxy         & $Type^{\dagger}$        & $(m-M)^{\ddagger}$ & $A_B^a$ &  Z & $12+log(O/H)$ \\ \hline
			NGC~5236 & SAB(s)c  & 27.84 & 0.29 & 0.008,0.019$^1$ & 8.2-8.6$^1$ \\ 
			NGC~7793 & SA(s)d  & 27.6 & 0.08 &  0.008,0.019$^2$ & 8.57$^5$  \\ 
			NGC~1313 & SB(s)d  & 28.2 & 0.47 & 0.004,0.008$^3$ & 8.33$^5$  \\ 
			NGC~45     & SA(s)dm  & 28.42 & 0.09 & 0.004,0.008$^4$ & ---  \\ \hline
%			NGC~4395 & SA(s)m  & 28.1 & 0.15$^3$ & 8.33$^2$ \\ \hline \hline
		\end{tabular} \\
	\label{tab:setparam}
	\flushleft{$^{\dagger}$  NASA/IPAC Extragalactic Database (NED); $^{\ddagger}$ \citet{mora09} and references therein; $^a$ \citet{schlegel98}; 
			     $^1$ \citet{bresolin09}; $^2$ \citet{calzetti10};$^3$ \citet{walsh97,larsen07}; $^4$ \citet{mora07}; and $^5$\citet{zaritsky94} at $r=3$ Kpc.}
\end{table*}

\begin{table*}[!t]
	\centering
	\caption{Journal of the observations.}
		\begin{tabular} {c c c c c c c c c} 
			\hline \hline
			Galaxy         & Number & F336W(U) & F435W(B) & F555W(V) & F814W(I) & RA. & DEC. & Date\\ 
			                      & of field & sec. & sec. & sec. & S & (J2000) & (J2000) & \\ \hline
			NGC~5236 & 1 & 2400 & 680 & 680 & 430 & 13:37:00 & -29:49:38 & 2004.07.28 \\ 
			                      & 2 & 2400 & 680 & 680 & 430 & 13:37:06 & -29:55:28 & 2004.08.07 \\ 
			NGC~7793 & 1 & 2400 & 680 & 680 & 430 & 23:57:41 & -32:35:20 & 2003.12.10 \\ 
			                      & 2 & 2400 & 680 & 680 & 430 & 23:58:04 & -32:36:10 & 2003.12.10 \\ 
			NGC~1313 & 1 & 2800 & 680 & 680 & 676 & 03:18:04 & -66:28:23 & 2004.07.17 \\ 
			                      & 2 & 2800 & 680 & 680 & 676 & 03:18:17 & -66:31:50 & 2004.12.18 \\ 
			                      & 3 & 2800 & 680 & 680 & 676 & 03:17:43 & -66:30:40 & 2004.05.27 \\ 
			NGC~45      & 1 & 2400 & 680 & 680 & 430 & 00:14:14 & -23:12:29 & 2004.07.05 \\ 
			                      & 2 & 2400 & 680 & 680 & 430 & 00:14:00 & -23:10:04 & 2004.06.01 \\  \hline 
		\end{tabular}
	\label{tab:journal}
\end{table*}

The paper is structured as follows. In Sect. 2, we present a short overview of previous work on 
our target galaxies, related to the present study. 
The basic reduction and characteristics of the observations are described
in Sect. 3. In Sect. 4 we present the photometry procedures applied to the data and describe how 
completeness tests were carried out. We also discuss the effect of stochastic sampling of
the stellar IMF on integrated cluster properties. In Sect. 5 we present the results of the estimation 
of ages and masses of clusters, as well as the field star formation histories. We also
estimate the cluster formation rates and use these to determine $\Gamma$ values.
In Sect. 6 we discuss our results and finally, we summarize and conclude our work in Sect. 7.

\section{Dataset overview}

In this paper we describe results for the remaining four galaxies in our HST/ACS sample:
NGC~5236, NGC~7793, NGC~1313, and NGC~45. These four galaxies share the properties of being 
face-on, nearby spirals; however, they differ in their morphology, star, and cluster formation 
histories. We present the basic properties of each galaxy in Table \ref{tab:setparam}.

\citet{LR99} studied cluster populations in a set of 21 galaxies, including the four included here. 
Using ground-based multiband ($UBVRI$ and H$\alpha$) observations
they estimated the total number of young massive clusters in each galaxy, using a 
magnitude limit of $M_V\le-8.5$. In a further work, \citet{LR00} estimated the star formation rate 
density ($\Sigma_{\rm SFR}$) and the specific $U$-band luminosity, 
$T_L(U) = 100\times$L(clusters,U)/L(galaxy,U), for each galaxy. The $T_L(U)$ was found to
correlate with $\Sigma_{\rm SFR}$. Taking $T_L(U)$ as a proxy for the 
cluster formation efficiency, these data thus suggested an increase in the cluster formation 
efficiency with $\Sigma_{\rm SFR}$.  
It is worth noting here that the $\Sigma_{\rm SFR}$ values were derived by normalizing the
total star formation rates, obtained from IRAS far-infrared fluxes, to the optical galaxy
diameters obtained from the RC3 catalog. Therefore, while these numbers were useful for studying
trends and correlations, they should not be taken as reliable absolute values.

More recent estimates of $\Sigma_{\rm SFR}$ have been made by \citet{calzetti10}
for the galaxies NGC~5236 and NGC~7793, where they found similar values to \citet{LR00}. 
\citet{harris01} present a photometric observation of clusters in the center (inner 300 pc.) 
of NGC~5236. Harris et al. find a large number
of young and massive clusters, consistent with a burst of star formation that began around 10 Myr ago,
but note that the apparent absence of older clusters might also be due to rapid disruption.
\citet{chandarwhitmore10} used the new Wide Field Camera 3 (WFC3) on HST to analyze the cluster 
system of NGC~5236. They find that luminosity functions and age distributions are consistent 
with previous work on 
galaxies of different morphological types \citep[e.g.][]{fall04}.
\citet{mora07,mora09} studied the cluster system for the same set of galaxies
used in this work, based on the same HST images. They present detailed estimates of the 
sizes, ages, and masses for the clusters detected. 
Mora et al. conclude that the age distributions are consistent with a $\sim$80\% MID per decade in age up to 1 Gyr,
 but could not make a distinction between different models (MDD vs.\ MID) of cluster
disruption. In the galaxy NGC~45 they found a large number of old globular clusters, of which
8 were spectroscopically confirmed to be ancient and metal-poor \citep{mora08}.

\section{Observation and data reduction}

The five galaxies studied in this series of papers were selected for detailed observations 
with the {\em Advanced Camera for Surveys} (ACS) and 
{\em Wide Field Planetary Camera 2} (WPFC2) onboard HST from the work of
\citet{LR99,LR00}. The two instruments have a resolution of $0\farcs05$ and $0\farcs046,0\farcs1$
for ACS and WFPC2 (PC,WFs), respectively. At the distance of our galaxies ($\sim4$ Mpc)
the ACS pixel scale corresponds to $\sim1$ pc.

Besides NGC~1313, which has three different fields observed, the rest of the galaxies were covered using two
pointings (see Fig. \ref{fig:set}). The bands used for the observations were F336W($\sim U$), F435W($\sim B$), 
F555W($\sim V$), and F814W($\sim I$), with the exposure times listed in Table \ref{tab:journal}.
The standard STScI pipeline was used for the initial data processing. ACS images were drizzled using the 
multidrizzle task \citep{koekemoer02} in the STSDAS package in IRAF using the default parameters, but 
disabling the automatic sky subtraction. WFPC2 images were combined and corrected for cosmic rays using the {\it crrej}
task using the default parameters. 

Object detection for field stars and star clusters was performed on an average B, V, and I image, using {\em daofind} 
in IRAF for the stars and SExtractor 
V2.5.0 \citep{bertinarnouts96} for the clusters. Coordinate transformations between ACS and WFPC2 used IRAF.
For details we refer to Paper I.

\section{Photometry}

We review here the procedures for carrying out photometry on our data, however, for details,
we refer  to Paper I.
Due to the crowding we performed PSF photometry for
field stars, while we used aperture photometry for the star clusters.

\subsection{Field stars}

With a set of bona-fide stars visually selected in our images, measuring their FWHM with {\it imexam}, we constructed
our point-spread function (PSF) using the PSF task in DAOPHOT. This procedure was followed in the same manner for each 
band (i.e., B,V, and I). The PSF stars were selected individually in each band, in order to appear bright
and isolated. PSF photometry was done with DAOPHOT in IRAF. 

HST zeropoints\footnote{www.stsci.edu/hst/acs/analisys/zeropints/\#tablestart} were applied to the PSF magnitudes 
after applying aperture corrections (see Sect. 4.3). The zeropoints used in this work are 
$ZP_B=25.767$, $ZP_V=25.727$ and $ZP_I=25.520$ magnitudes. Typical errors of our photometry do not change 
dramatically from the ones in Paper I (see its Fig. 2). 

\begin{figure}[!t]
	\centering
		\includegraphics[trim= 0mm 0mm 0mm 0mm,width=\columnwidth]{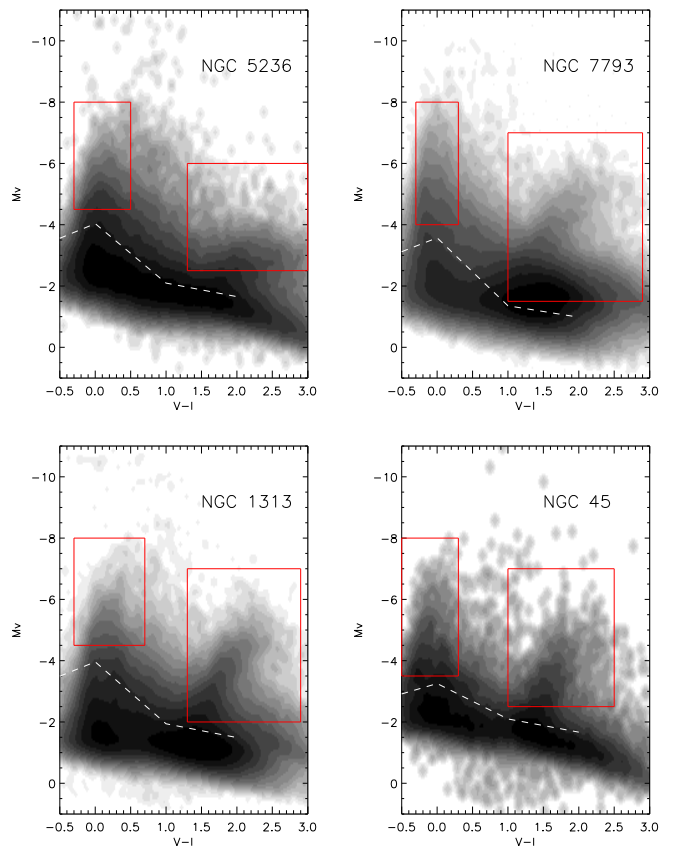}
		\caption{Hess diagram for the field stars of the observed galaxies. The dashed white line 
	         	represents the $50\%$ completeness curve. Red lines enclose the fitted areas used to estimate 
	         	the SFH (see Sect. 5).}
	\label{fig:hess}
\end{figure}

Having magnitudes for our field stars, Hess diagrams were constructed and are depicted in Fig. \ref{fig:hess}
(each panel presents all fields combined for each galaxy). The total number of stars varies among the galaxies, 
all having some tens of thousands. 
Various phases of stellar evolution can be identified in the Hess diagrams:

\begin{itemize}
\item main sequence and possible blue He-core burning  stars at $V-I \sim0$ and $-2\le V\le-8$;
\item red He core burning stars at $1.2\le V-I \le2.5$ and $-2.5\le V\le-6.5$; 
\item RGB/AGB stars, near the detection limit at $1\le V-I \le3$ and $-0.5\le V\le-2.5$.
\end{itemize}

The same features were observed for NGC~4395 in Paper I.

Overplotted in Fig. \ref{fig:hess} are the $50\%$ completeness lines (see Sect. 4.5 for details of the completeness analysis).
Also, red lines enclose the fitted areas that will be used in Sect. 5 to estimate the star formation histories of the galaxies.
These areas were selected to cover regions that were clearly over the 50\% completeness and represent stars in different stages of
evolution (e.g. main sequence, red He core burning).

\subsection{Star clusters}

To detect the cluster candidates we used SExtractor with a detection criterion of six connected pixels and
a threshold of 10 sigma above the background level. The total numbers of objects detected in each galaxy are listed 
in the second column of Table \ref{tab:clustersdetected}. For these objects aperture photometry was performed
using an aperture radius of six pixels on our ACS pointings, corresponding to about two half-light radii 
for a typical star cluster. We used a sky annulus with five pixels width and an inner radius of eight pixels. For the 
WFPC2 images, the apertures used cover the same area. There is a possibility of having close-neighbor objects 
that contaminate the photometry, whether inside either of the aperture radii or the sky annulus. Sizes
were measured using the ISHAPE task in the BAOLAB package \citep{larsen99}. 

As mentioned in Paper I, three criteria were used to produce catalogs of cluster candidates from
the initial SExtractor output.

\begin{enumerate}
\item {\it Size}: Candidates must satisfy $FWHM_{\rm SExtractor} \geq2.7$ pixels and $FWHM_{\rm ishape} \geq 0.7$ pixels. 
These are rather conservative size cuts that may eliminate some of the most compact clusters, but reduce the risk
of contamination from other sources.
\item {\it Color}: Candidates must satisfy  $V-I \le1.5$. 
\item {\it Magnitude}: Candidates must be brighter than $m_V=23$ ($M_V$ brighter than $-4.6$ to $-5.4$, depending on the galaxy distance).
\end{enumerate}
\begin{table*}[!t]
	\centering
	\caption{Number of clusters detected per field, per galaxy.
	   	     Clusters with measured sizes (2$^{nd}$ column), total of objects with three band photometry(3$^{rd}$ column),
		    accepted (4$^{th}$ and 5$^{th}$ columns), suspected (6$^{th}$ and 7$^{th}$ columns), and total rejected (8$^{th}$ column). Subscripts 3B and 4B represent 
		    three and four band photometry. Shaded areas indicate the total per galaxy.}
		\begin{tabular} {c c c c c c c c} 
			\hline \hline
			 Galaxy$\_$F & Ishape & T$_{3B}$ & A$_{3B}$ & A$_{4B}$ & S$_{3B}$ & S$_{4B}$ & R$_{total}$  \\ \hline
			NGC 5236$\_$F1 & 9788 & 1027 & 286 & 117 & 519 & 255 & 222 \\ 
			NGC 5236$\_$F2 & 7290 & 758 & 274 & 85  & 326 & 123 & 158 \\ 
\rowcolor{lightgray}	NGC 5236   & 17078 & 1785 & 560 & 202 & 845 & 378 & 380 \\ \hline 
			NGC 7793$\_$F1 & 12095 & 521 & 83 & 41 & 308 & 150 & 130 \\ 
			NGC 7793$\_$F2 & 13597 & 274 & 72 & 34   & 95 & 24 & 107 \\ 
\rowcolor{lightgray}	NGC 7793   & 25692 & 795 & 155 & 75 & 403 & 174 & 237 \\ \hline 
			NGC 1313$\_$F1 & 19925 & 1033 & 184 & 70 & 288 & 79 & 561\\ 
			NGC 1313$\_$F2 & 13153 & 751 & 164 & 52   & 115 & 15   & 472 \\ 
			NGC 1313$\_$F3 & 12287 & 133 & 57 & 28   & 7 & 2   & 69 \\ 
\rowcolor{lightgray}	NGC 1313   & 45365  & 1917 & 405 & 150 & 410 & 96 & 1102 \\ \hline 
			NGC 45$\_$F1 & 3760 & 46 & 22 & 12 & 2 & 1 & 22 \\ 
			NGC 45$\_$F2 & 4634 & 92 & 45 & 23 & 11 & 5   & 36 \\ 
\rowcolor{lightgray}	NGC 45   & 8394 & 138 & 67 & 35 & 13 & 6 & 58 \\ \hline 			
			
		\end{tabular}
	\label{tab:clustersdetected}
\end{table*}

\begin{figure}[!t]
	\centering
		\includegraphics[width=\columnwidth]{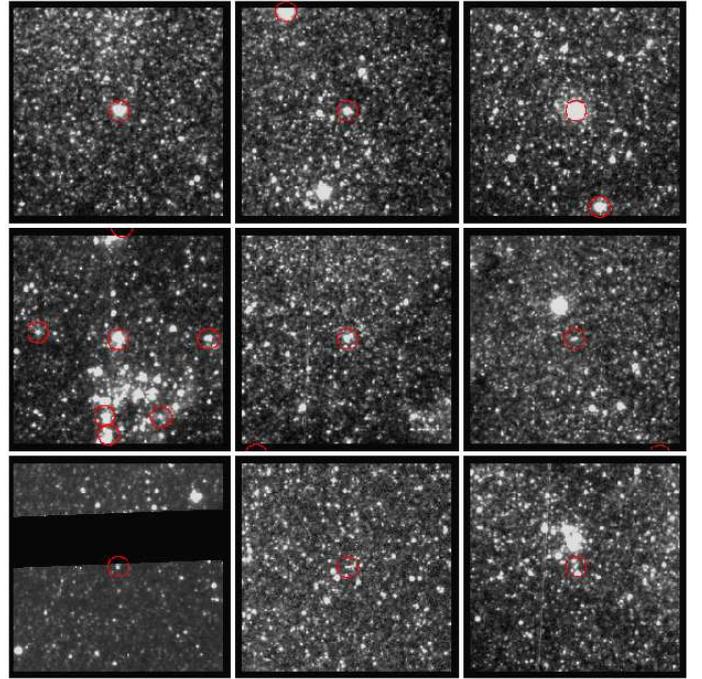}
		\caption{Examples of objects that were accepted (first row), suspected (second row), and rejected (third row) 
		 after visual inspection. Stamps are from the first field observed in NGC~7793 and have sizes of 100$\times$100 pixels.}
	\label{fig:rejects}
\end{figure}

Since the WFPC2 fields only cover about half the area of the ACS fields, some objects will only have three-band photometry ($BVI$), while
others will have all four colors.
Objects that satisfy the three criteria listed above are considered as star cluster candidates in the rest of
the paper. However, as found in many previous studies, there is no unique combination of objective
criteria that can lead to a successful detection of bona-fide clusters and no false detections.
Our cluster candidates were therefore visually inspected to determine whether they resemble star clusters. 
Based on this, we classified the cluster candidates into three
categories: Accepted, Suspected, and Rejected. Figure \ref{fig:rejects} presents some examples of each
category. In this figure, the first row
presents the Accepted objects, which are clearly extended objects with normal measured sizes and magnitudes. 
The second row presents the Suspected objects, where the size/magnitude measurements may be affected by
crowding, where the shape appears irregular, or where the contrast against the background is not strong.
The last (third) row presents examples of the Rejected objects.

Table \ref{tab:clustersdetected} summarizes the total number of objects detected that have size 
measurements ($2^{nd}$ column), the total number of objects with three-band photometry
and have sizes over the limits imposed ($3^{rd}$ column),
the total number of accepted objects with three- and four band photometry 
($4^{th}$ and $5^{th}$ columns), the total number of suspected objects with three- and four band 
photometry ($6^{th}$ and $7^{th}$ columns), and the total number of rejected objects ($8^{th}$ column). 
Shaded areas are the total numbers per galaxy.
 
\begin{figure}[!t]
	\centering
		\includegraphics[trim= 0mm 0mm 0mm 0mm,width=\columnwidth]{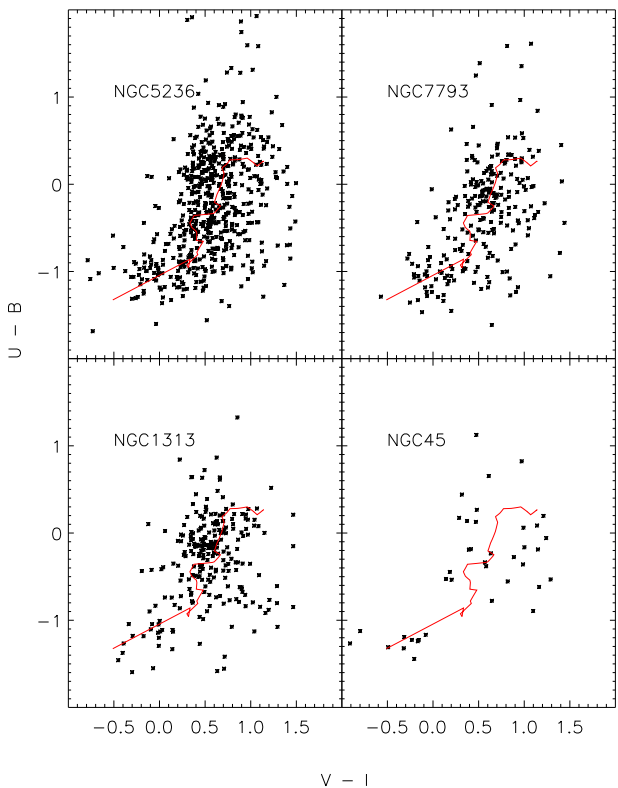}
		\caption{Two-color diagrams for the clusters with four band photometry. The red line represents 
		the theoretical path a cluster
		will follow using GALEV models, assuming LMC metallicity and no extinction. Accepted+ Suspected clusters are presented.}
	\label{fig:twoc-obs}
\end{figure}

Figure \ref{fig:twoc-obs} shows two-color diagrams for accepted plus suspected clusters with four band photometry 
(all the fields combined per galaxy),
corrected for foreground extinction with the values presented in Table \ref{tab:setparam}. Overplotted  is
a theoretical track  that a cluster will follow between 4 Myr and 1 Gyr using Galev models \citep{andersfritze03},
assuming LMC metallicity and no extinction. We see that the clusters generally tend to align with the
model sequence, but with significant scatter around it. Below we investigate to what extent this scatter
may come from stochastic color variations due to random sampling of the stellar IMF.

\subsection{Completeness}
Completeness analysis was carried out separately for field stars and star clusters to account
for both populations.

\subsubsection{Field stars}
As in Paper I, we created artificial stars using the PSF obtained in Sect. 4.1.
In the magnitude range between 20 to 28, every 0.5 magnitudes, we generated
5 images and passed each one through the photometry procedures, using the exact same
parameters as are used for the original photometry. A total of 528 stars were added to
each image, with a separation of 100 pixels (we did not take subpixels shifts into account). 
The images were created using
{\it mksynth} in BAOLab \citep{larsen99} and added to the science images using
{\it imarith} in IRAF. To quantify the dependency of completeness functions on color, we
made use of the near 1:1 relation between the $B-V$ and $V-I$ colors of stars 
(see Paper I for details).

Based on our analysis, we found $50\%$ completeness limits for each galaxy and for each band,
as shown in Table \ref{tab:cmplset_fs}. 
%The color dependencies
%are depicted as dashed (white) lines in Fig. \ref{fig:hess}.

\begin{table}[!t]
	\centering
	\caption{50 \% completeness limits for field stars.}
		\begin{tabular} {c c c c c} 
			\hline \hline
			 Galaxy & Field & F435W(B) & F555W(V) & F814W(I) \\ \hline
			NGC~5236 & 1 & 26.12 & 26.10 & 25.10 \\ 
			NGC~5236 & 2 & 26.48 & 26.35 & 25.48 \\ 
			NGC~7793 & 1 & 26.60 & 26.52 & 25.25 \\ 
			NGC~7793 & 2 & 26.91 & 26.78 & 26.03 \\ 
			NGC~1313 & 1 & 26.64 & 26.66 & 26.15 \\ 
			NGC~1313 & 2 & 26.65 & 26.55 & 26.14 \\ 
			NGC~1313 & 3 & 26.79 & 26.78 & 26.41 \\ 
			NGC~45 & 1 & 26.77 & 26.74 & 26.28 \\
			NGC~45 & 2 & 26.67 & 26.60 & 26.08 \\  \hline 
		\end{tabular}
	\label{tab:cmplset_fs}
\end{table}

\subsubsection{Star clusters}
\begin{figure}[!t]
	\centering
		\includegraphics[trim= 0mm 0mm 0mm 0mm,width=\columnwidth]{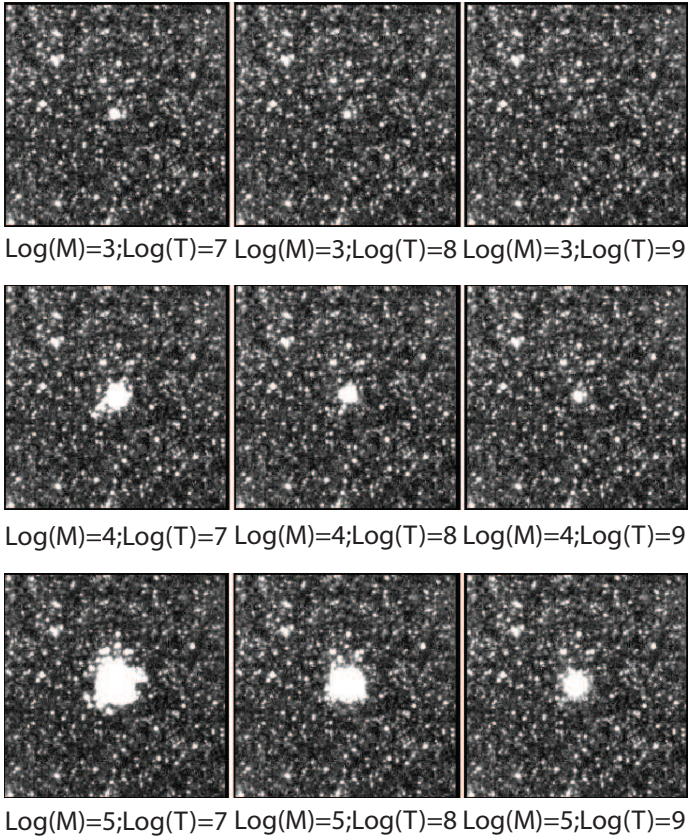}
		\caption{Stochastic clusters created to estimate the completeness in NGC~7793 
		         (average B, V, and I image, first field). 
			     From top to bottom, rows represent masses of log(M)[M$_{\odot}$]=[3,4,5]. 
			     From left to right, columns represent ages of log($\tau$)[yr]=[7,8,9]. Each images has a size of $100\times100$ pixels.}
	\label{fig:clus_im}
\end{figure}

In order to quantify completeness limits we added artificial clusters
of different ages and masses to our images. We created artificial clusters using a
stochastic approach. Assuming a Kroupa IMF \citep{kroupa02} in the mass range 0.01 to 100 M$_{\odot}$, 
a total cluster mass of  $M=[10^3,10^4,10^5]$ M$_{\odot}$, and a cluster
age range between $\tau=[10^7,10^{9.5}]$ yr (with 0.5 dex steps), 
we randomly sampled stars from the IMF until the total mass of the 
stars reached the total mass assumed for the cluster. Positions were assigned by randomly
sampling a King profile \citep{king62}.
There is a possible pitfall regarding the mass of the last star sampled, because it could
overcome the total input (assumed) mass. We kept the last star, even if the total mass is higher than assumed. 
This problem affects low-mass clusters more
than high-mass clusters. With the ages and masses for the stars, we then interpolated in isochrones (of LMC-like
metallicity) from
the Padova group \citep{marigo08} and assigned magnitudes
to each star. For all the artificial clusters, an $FWHM=2.7$ pixels was assumed
(corresponding to a $R_{\rm eff}\approx4$ pc). 
Figure \ref{fig:clus_im} shows stamps of artificial clusters of different ages and masses,
using an average (B, V, and I) image of the galaxy NGC~7793 as an example.

For each combination of age and mass, a total of 100 randomly generated clusters were added to the
science images using a square grid. The artificial images with clusters were created using {\it mksynth} 
in BAOLab \citep{larsen99} and added to the science images using {\it imarith} in IRAF. Following the same 
procedure used for the cluster
photometry, an average $BVI$ image was created for each field.  SExtractor was then run on this
average image, using the same parameters
as in Sect. 4.2. SExtractor returns a file with coordinates, measured FWHM and other information. 
To save computational time,
owing to the large amount of objects that SExtractor could detect, we removed all the original 
objects (science objects detected previously) from the list and kept
the ones that are not in the original image. This new coordinate file was passed
to {\it ishape} in BAOLab to compute PSF-corrected sizes. We used the 
coordinate file to run photometry, again with the same procedures and parameters as for
the science photometry. Having B, V, I photometry done, size cuts were applied and 
the output file was matched with the input coordinate file to evaluate how many
of the added artificial objects were recovered successfully. 

\begin{figure}[!t]
	\centering
		\includegraphics[trim= 0mm 0mm 0mm 0mm,width=\columnwidth]{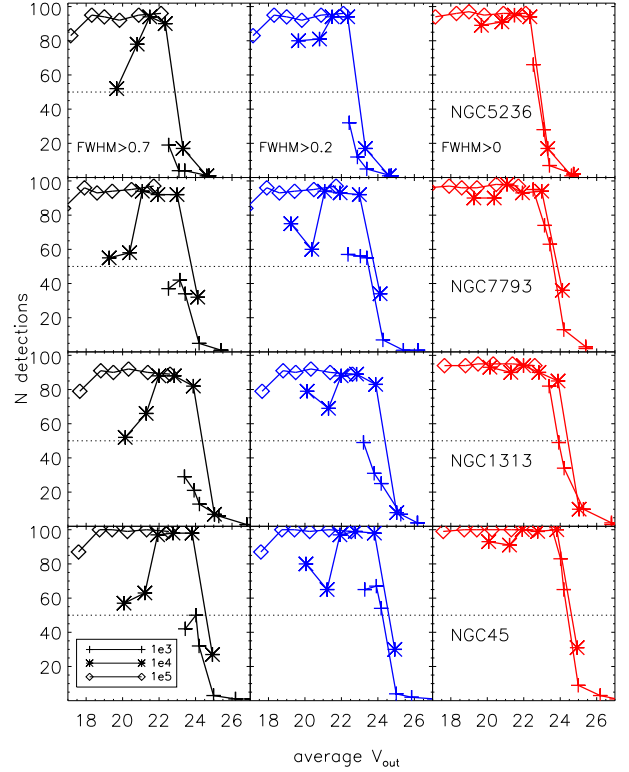}
		\caption{Completeness curves for stochastic clusters with ages log($\tau$)[yr]=[7,7.5,8,8.5,9,9.5] and masses of log(M)[M$_{\odot}$]=[3,4,5] for
			      the four galaxies. All clusters have an input FWHM=2.7 pixels.
			      Red lines represent the number of detections without applying any size criteria (right column). 
			      Blue lines are the number of detections after applying the size criteria of $FWHM_{ishape}\ge0.2$ pixels, used in \citet{mora07,mora09} (middle column).
			      Black lines are the number of detections after applying the size criteria used in this work ($FWHM_{ishape}\ge0.7$ pixels) (left column). 
			      The legend is in units of solar masses (M$_{\odot}$). The symbols over the lines
			      represent an age step (i.e. 0.5 dex) starting from the left.}
	\label{fig:cmpl_clu}
\end{figure}

Figure \ref{fig:cmpl_clu} shows the output ($m_V$) average magnitude of the recovered clusters versus the fraction 
recovered for the three masses and six ages assumed in each galaxy. 
For the masses log(M)[M$_{\odot}$]=[3,4,5] as an illustrative example, Fig. \ref{fig:cmpl_clu} shows the total 
number of objects detected applying three different size criteria: (1.) $FWHM_{ishape}>0$, i.e.\ no size cut used to 
select the clusters; (2.) $FWHM_{ishape}\ge0.2$, same size cut used by \citet{mora07,mora09}; and (3.) $FWHM_{ishape}\ge0.7$,
the size cut used in this paper. Each line is for a given cluster mass, while each symbol belonging 
to a line represents the time steps assumed, i.e., $\tau=[10^7,10^{9.5}]$ yr (with 0.5 dex step). 
From this test we conclude that

\begin{enumerate}
\item High-mass clusters (log(M)[M$_{\odot}$]=5), at any age, are easily recognized by our procedures, regardless of
the size criteria used. 
\item For decreasing cluster mass, clusters of old ages drop out of the sample, and for log($\tau$)[yr]$\ge$8.5 our completeness is less than 50\% for masses below log(M)[M$_{\odot}$]=4. 
\item The completeness depends on the mass, on the size used to 
	classify an object as extended or not extended, and on age.
	Decreasing the size threshold would increase the completeness somewhat for
	low-mass, young objects, but could
	introduce additional contamination.
\item Stochasticity affects young star clusters (log($\tau$)[yr]$\le$7.5) more dramatically.
\end{enumerate}

\begin{figure}[!t]
	\centering
		\includegraphics[trim= 0mm 0mm 0mm 0mm,width=\columnwidth]{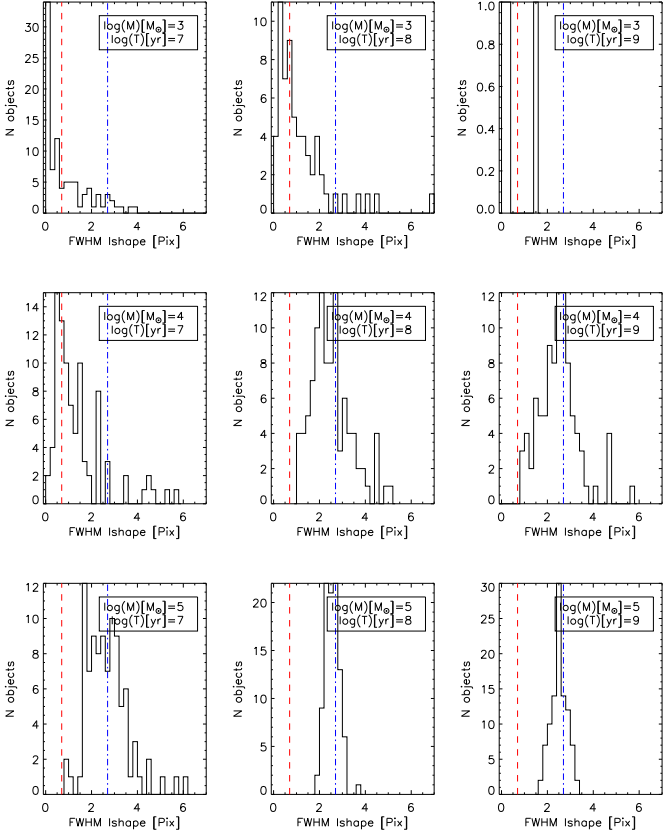}
		\caption{Histograms of measured FWHMs for the stochastic clusters in NGC~7793 
		         created for the completeness analysis. From top to bottom, rows 
			     represent masses of log(M)[M$_{\odot}$]=[3,4,5]. From left to right, 
			     columns represent ages of log($\tau$)[yr]=[7,8,9].
			     Legends are in logarithmic mass and age units. The red dashed line represents an FWHM=0.7 pixels, as assumed in Sect. 4.3.2, while the blue 
			     dash-dotted line represents the input FWHM=2.7 pixels.}
	\label{fig:clu_fwhms}
\end{figure}

At the very young ages in Fig. \ref{fig:cmpl_clu} the completeness drops below 100\% at all masses. 
This can be understood from Fig. \ref{fig:clu_fwhms}, which shows the histograms for the FWHM
of the detected objects measured with {\it ishape} for log(M)[M$_{\odot}$]=[3,4,5] and log($\tau$)[yr]=[7,8,9], 
using the results from NGC 7793 as an example. 
%The blue dotted-dashed lines represent the input sizes, while the red
%dashed lines indicate our size cut. 
The figure shows that at high masses and old ages, the recovered
sizes are on average similar to the input values, with some spread. However, at younger ages and/or lower
masses, the measured sizes are systematically less than the input values. In these cases, the light
profiles can be dominated by a single or a few bright stars, while for high masses and/or old ages, 
the light profiles are much smoother and better fit by the assumed analytic profiles. At young ages
and low masses, this bias in the size measurements leads to a decrease in the completeness fraction as
more clusters fall below the size cut.

\subsection{Aperture corrections}

Aperture corrections were estimated separately for star and clusters. We applied the same procedures as in 
Paper I.  For field stars, our PSF-fitting magnitudes were corrected to a nominal aperture radius of $0\farcs5$, following 
standard procedures. From this nominal value to infinity, we applied the corrections in \citet{sirianni05}.

For star clusters, aperture corrections were applied following the equations in \citet{mora09}, which give
a relation between the FWHM of the objects and the aperture corrections. The photometric parameters (and data set)
used in our work are the same as the ones used by \citet{mora09}, allowing us to use their equations. This set of equations
apply corrections to a nominal radius of $1\farcs45$. We adopted the values in \citet{sirianni05} to correct from there on,
although the corrections to infinity are minor ($\sim97\%$ of the total energy is encircled within $1\farcs5$).

\subsection{How do stochastic effects influence star cluster photometry?}

\begin{figure*}[!th]
	\centering
		\includegraphics[scale=0.85]{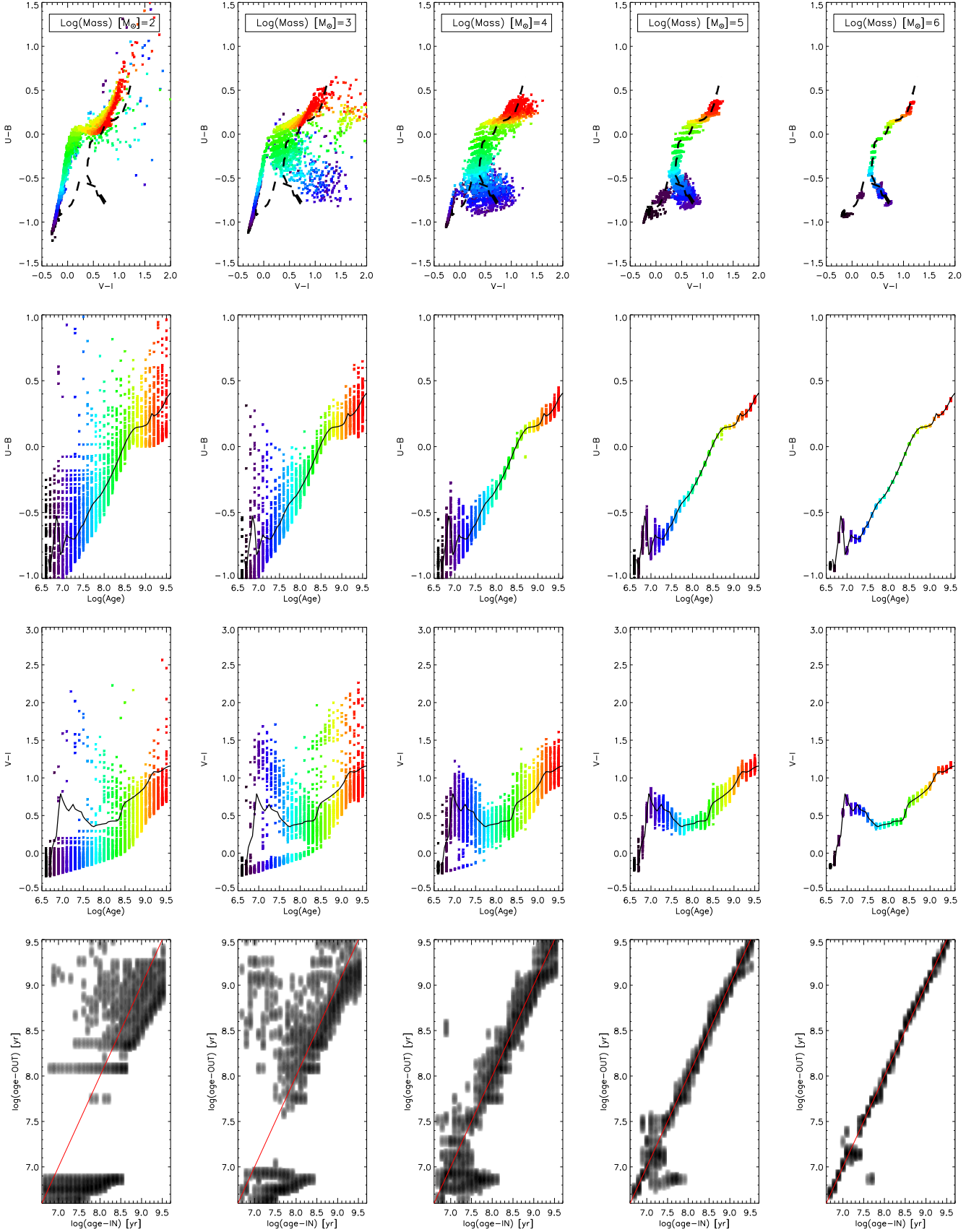}
		\caption{Stochastic effects on colors and ages of clusters. {\em First row}:Two color diagrams for 
		         a stochastic sample of clusters 
			     with different masses and different ages. The dashed line represents  Padova 2008 SSP models of
			     solar-like metallicity. 
			     {\em Second row}: U-B color 
			     evolution. {\em Third row}: V-I color evolution. {\em Fourth row}: Density plot to make a comparison between input and output ages. The red line represents the 1:1 relation,  
			     not a fit to the data.}
	\label{fig:stoch2c}
\end{figure*}

The classical approach used to estimate masses and ages of unresolved star clusters
in the extragalactic field is based on the use of multi-band integrated colors and
comparison of these with SSP models. 
\citet{anders04} show that it is necessary to have at least four-band photometry to be able to
break degeneracies (e.g. age-metallicity). However, standard SSP models assume a continuously 
populated stellar initial mass function (SIMF), while clusters consist of a finite number of
stars.  
For unresolved clusters, the random sampling of the SIMF can strongly affect integrated properties
such as cluster colors, magnitudes, and parameters (ages, masses) derived from them
\citep{cervignoluridiana06,maizapellaniz09,popescu10a,popescu10}. A promising attempt to take the
stochastic color fluctuations into account when deriving ages and masses has been made by
\citet{fouesneau10}, based on a Bayesian approach. 

Here we do not attempt to offer any solution to the SIMF sampling problem, but we quantify
its effects that are related to our study. To that aim, we 
created clusters by randomly sampling the SIMF and assigning magnitudes to individual stars in the
same bands used for our photometry (i.e. U, B, V, and I). 
In addition to its impact on cluster detection and classification, as described in the previous
section, we also investigated how stochastic SIMF sampling affects the two-color diagrams and ages.

We used the same recipe as in Sect. 4.3.2 to create artificial clusters. Assuming a range
of ages between $10^{6.6}$ Myr and $10^{9.5}$ yr and total masses of M=[$10^2,10^3,10^4,10^5,10^6$] M$_\odot$,
we created 100 clusters every 0.1 dex in age. The top row in Fig. \ref{fig:stoch2c} shows the two-color diagrams 
for each one of the total masses, together with an solar-metallicity Padova SSP model \citep{marigo08}. The evolution of the
colors U-B and V-I with time are shown in the second and third rows, 
and the comparison of input (assumed) and output (estimated) ages using AnalySED \citep{anders04} are in the 
bottom row. The colors indicate the input ages. 

Many features are observed here.
(1.) For high masses (log(Mass)[M$_{\odot}$]=[5,6]), the stochastically sampled clusters form narrow sequences
in the color-color and color-age diagrams, in agreement with \citet{cervignoluridiana06}. 
(2.) At intermediate masses, i.e., log(Mass)[M$_{\odot}$]=[3,4], which are typical of the cluster
masses observed in extragalactic works, the scatter increases strongly. The scatter observed in
the two-color diagrams for this mass range is similar to what is seen in our observed
two-color diagrams.
(3.) For very low masses (log(Mass)[M$_{\odot}$]=[2]), the scatter again decreases but the model colors
now deviate strongly from the SSP colors. This is because such clusters have a very low probability of
hosting a luminous (but rare and short-lived) post-main sequence star, while the SSP models assume that the 
colors are an average over all stages of stellar evolution \citep[see][]{piskunov09}. 
(4.) For ages log($\tau$)$\lesssim8$ and intermediate masses (log(Mass)[M$_{\odot}$]=[3,4]), the color distribution
actually becomes bimodal, as observed by \citet{popescu10a,popescu10}. The blue ``peak'' in the color distribution
is due to clusters without red supergiants, while the presence of even a single red supergiant shifts the
colors into the other peak.
(5.) The bottom row shows that age estimates are completely dominated by stochastic effects for low-mass clusters 
(log(Mass)[M$_{\odot}$]=[2,3]). For higher masses the
 ages are more accurately recovered, even if a small scatter is observed compared to the 1:1 relation, 
 especially at ages of a few tens of Myr where the light is strongly dominated by red supergiant stars.

Based on these results it is clear that photometry and the ages estimated from broad band photometry can be 
heavily affected by the stochastic effects introduced by SIMF sampling, as also shown by \citet{maizapellaniz09}. 
We conclude that low-mass clusters ($M \le 10^3 M_{\odot}$) are very strongly affected, reaching age differences up to 
log(age$_{in}$)-log(age$_{out})\approx2.5$ dex, for clusters with log($\tau$)[yr]$\le8.5$,
while for high-mass clusters ($M \ge 10^4 M_\odot $), this effect gradually diminishes, an effect that is clearly visible
as gaps in Fig. \ref{fig:stoch2c} (last row). Another important effect is observed as a deviation from the 1:1 red line observed
for the low-mass clusters, which indicate that the estimated age ($Age_{out}$) is again wrongly recovered, even at old ages (1 Gyr).
One should be aware of the risk that low-mass, young clusters may erroneously be assigned old ages.
When using SSP models to convert their luminosities to masses, based on such wrong age estimates, such
objects might be assigned erroneously high masses, thus making it into an observed sample
\citep[e.g.][]{popescu10}. 

It is important to mention that the SSP models used do not take the binarity or rotation of massive stars into account  \citep[see e.g.][]{maedermeynet08, eldridgestanway09}. Also, different
isochrone assumptions and the techniques used to perform the fit to the ages and masses can be affecting the results  
\citep[see e.g][]{scheepmaker09,degrijsanders05}.
The tests presented in this paper are only intended to address the stochastic sampling effects, and we only rely on Padova isochrones.
A more detailed study must be made to better account for other effects (e.g. by binaries, rotation, etc).

\begin{figure}[!ht]
	\centering
		\includegraphics[trim= 0mm 0mm 0mm 0mm,width=\columnwidth]{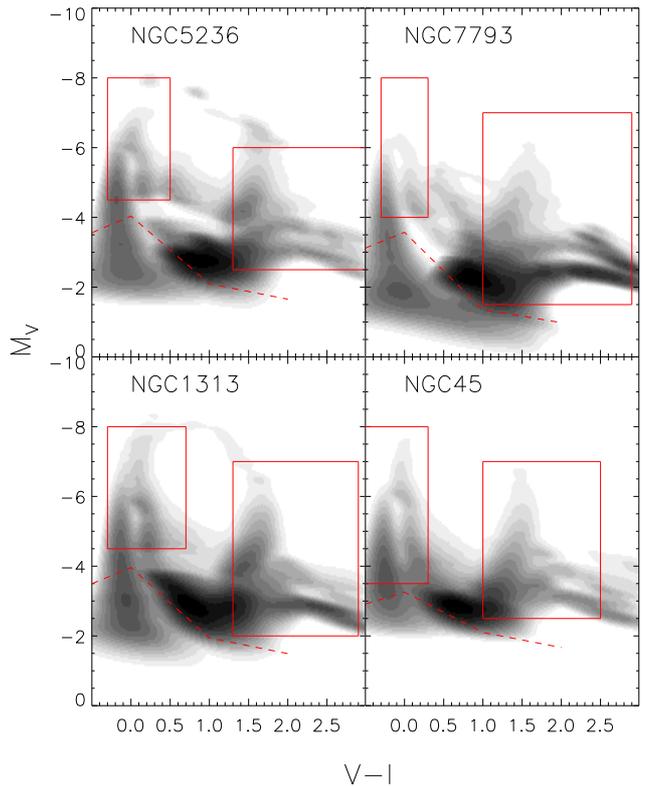}
		\caption{Fitted Hess diagrams for the sample studied. Red lines are the boxes used for the fit and the 50\% completeness, same as Fig. \ref{fig:hess}.}
	\label{fig:fit_sfhs}
\end{figure}

\begin{figure}[!t]
	\centering
		\includegraphics[trim= 0mm 0mm 0mm 0mm,width=\columnwidth]{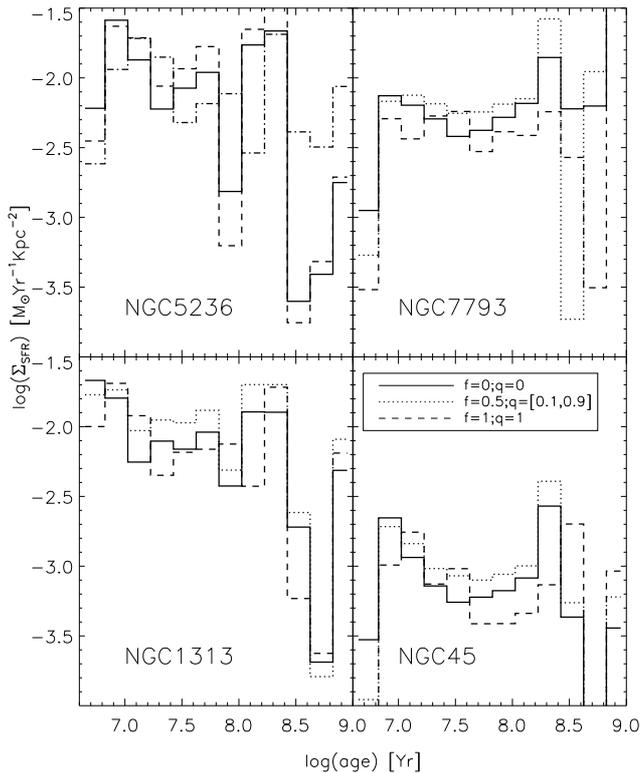}
		\caption{Star formation rate densities for the data set. Each line represent an assumption for the binarity (see text for details).}
	\label{fig:sfhs}
\end{figure}

\section{Results}

In the next sections we estimate the field star formation histories (SFHs) for each galaxy 
and determine the ages and masses of the cluster candidates. We then estimate the
cluster formation efficiencies, $\Gamma$.

\subsection{Field star formation histories}

To estimate the SFHs, we used the synthetic CMD method. We implemented
this method using an IDL-based program that was introduced and tested in Paper I. For
a description of the program we refer the reader to that paper, but here we summarize the basic functionality. 

The synthetic CMD method \citep{tosi91} takes advantage of the power supplied by the CMDs. The method uses 
a group of isochrones, together with assumptions about the SIMF, metallicity, distance, extinction, and binarity,
to reproduce an observed CMD. Photometric errors and completeness functions (both being treated as magnitude dependent parameters by our program) are
also taken into account. The program searches for the combination of isochrones that best matches the
observed CMD, thereby estimating the SFH. 

The parameters used to estimate the SFH of the galaxies are: 
a Hess diagram with a resolution of $200\times200$ pixels is created, using a Gaussian kernel 
with a standard deviation of 0.02 magnitudes along the color axis. The matching is done within
the rectangular boxes depicted in Fig. \ref{fig:hess}, using the V-I vs. V color combination, 
Padova 2008 isochrones \citep{marigo08} and a \citet{kroupa02} IMF in the mass range 0.1 to 100 M$_{\odot}$.
The assumed distance moduli, foreground extinctions, and metallicities are given in 
Table \ref{tab:setparam}, while the photometric errors and completeness for each galaxy were determined
in Sect. 4. The program also includes a simplified treatment of binaries, in which binary evolution is ignored,
but the effect of unresolved binaries on the CMD are modeled.
To account for binarity we used three different assumptions
for the binary fraction ($f$) and mass ratio ($q$): (1.) $f=0.0$ and $q=0.0$, (2.) $f=0.5$ and $q=[0.1,0.9]$ (assuming a flat distribution), and (3.) $f=1$ and $q=1$.
These three assumptions are the same ones as used in Paper I. 

Figures \ref{fig:fit_sfhs} shows the best-fit Hess diagrams. Comparing with the observed Hess diagrams
(Fig. \ref{fig:hess}), we see that the fits are far from perfect. In particular, all the model Hess 
diagrams show a clear separation between the blue core He burning (``blue loop'') stars and the
main sequence, while this is not obvious in most of the observed diagrams. This might be partly due
to some variation in the internal extinction, which has not been included in our modeling.
To infer the SFHs of our sample, we combined all the fields (per galaxy) and passed to our program. 
The star formation rates, normalized to unit area, are shown in Figure \ref{fig:sfhs} and the average
values are listed in Table \ref{tab:sfrcfr} for ages between 10 and 100 Myrs. In this age range,  
our data are less affected by incompleteness. Previous estimates of the $\Sigma_{\rm SFR}$ done 
by \citet{LR00} and  \citet{calzetti10} are included in table \ref{tab:sfrcfr}. 
We see that NGC~5236 and NGC~1313 have higher $\Sigma_{\rm SFR}$
values than NGC~7793 and NGC~45, in agreement with the previous estimates. 
Within the uncertainties  (see Paper I), we do not see any significant trends in the SFRs
between $10^7$ and $10^8$ years.
While our estimate of $\Sigma_{\rm SFR}$ agrees very well with the others for NGC~5236, there are 
significant differences
for some of the other galaxies, most notably for NGC~4395. It should be kept in mind that the
$\Sigma_{\rm SFR}$ values derived here are for our specific ACS fields, while the others are
averages over whole galaxies over a rather large outer diameter. It is therefore not very surprising
that our new estimates tend to be higher.

\begin{table*}[!th]\centering
\caption{Estimates of the star formation rates.}
\begin{tabular}{c c c c c c}
\hline \hline
Galaxy  & SFR$^{\dagger}$ &   Areas$^{\dagger}$ & $\Sigma_{sfr}^{\dagger}$ & $\Sigma_{sfr}^{\ddagger}$  & $\Sigma_{sfr}^{a}$ \\ 
  & [M$_{\odot}$yr$^{-1}]$  & [Kpc$^2$] & [M$_{\odot}$yr$^{-1}$Kpc$^{-2}$] & [M$_{\odot}$yr$^{-1}$Kpc$^{-2}$] &  [M$_{\odot}$yr$^{-1}$Kpc$^{-2}$] \\ \hline 
NGC5236  & 0.39  &    28.71         & $13.43\times10^{-3}$  &         $13.8\times10^{-3}$   &         $ 16.8\times10^{-3}$ \\
NGC7793  & 0.15  &    23.05    & $6.43\times10^{-3}$  &         $2.12\times10^{-3}$   &         $2.8\times10^{-3}$     \\
NGC1313  & 0.68  &    60    & $11.26\times10^{-3}$&        $4.04\times10^{-3}$   &           ---                       \\
NGC45      & 0.05  &   48.99      & $1.01\times10^{-3}$  &         $0.23\times10^{-3}$   &           ---                         \\ 
NGC4395$^b$  &  0.17 & 36.48 &  $4.66\times10^{-3}$  &         $0.25\times10^{-3}$ & ---  \\ \hline 
\end{tabular}
\label{tab:sfrcfr}
	\flushleft{$^{\dagger}$estimated in this paper; $^{\ddagger}$\citet{LR00}; $^a$\citet{calzetti10}; $^b$ values from Paper I.}
\end{table*}

\subsection{Cluster ages and masses}
To determine the ages and masses of the clusters we used the program AnalySED \citep{anders04}. Using GALEV SSP models \citep{schulz02},
AnalySED compares the observed spectral energy distributions with a library of models to find the best fit. We
used GALEV models based on a Kroupa IMF \citep{kroupa02} in the mass range 0.1 to 100 M$_{\odot}$, Padova 
isochrones \citep{girardi02}, and different metallicities, depending on the galaxy.

\begin{figure}[!t]
	\centering
		\includegraphics[trim= 0mm 0mm 0mm 0mm,width=\columnwidth]{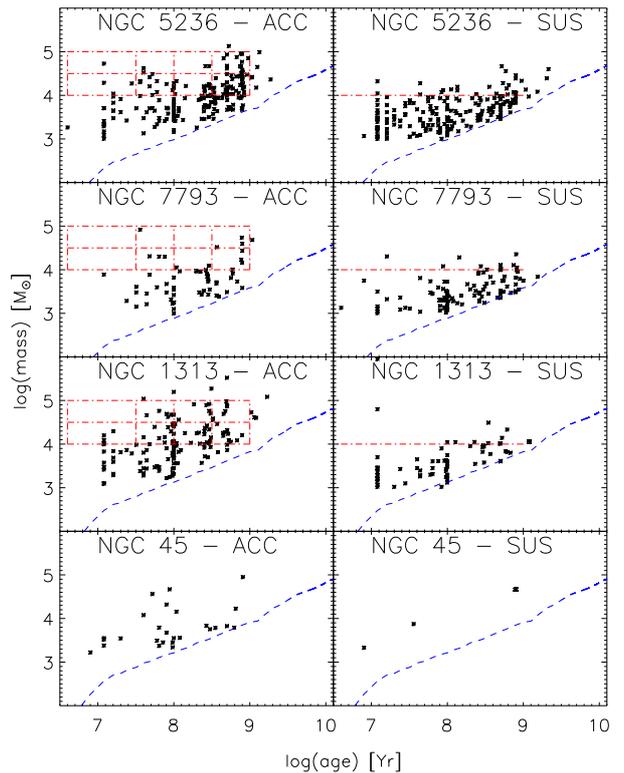}
		\caption{Age-mass distributions for the cluster systems. Blue dashed lines represent the magnitude 
		             cut at $m_v=23$. Red boxes will be used to create
			    the age and mass distributions. Left column present the distribution for the Accepted sample, while the right column present
			    the suspected sample of clusters (see Sect. 4.2). Red dash-dotted line in the right column denotes the mass 10$^4$ M$_{\odot}$ for comparison.}
	\label{fig:agemass}
\end{figure}

Based on our analysis of uncertainties due to stochasticity, we applied a mass criterion to the cluster samples in
addition to the three selection criteria defined in Sect. 4. We require clusters in our sample
to be more massive than 1000 M$_{\odot}$; however, we remind the reader that ages are only reliable
for masses greater than $10^4$ M$_{\odot}$ (for ages $\tau\le$1 Gyr). The magnitude limit, 
$M_V\sim-5$, is generally below our 50\% detection limit based on the stochasticity test.

Figure \ref{fig:agemass} shows the age-mass diagrams for clusters that satisfy the four criteria for the accepted and suspected samples separated. 
Overall, we observe that the suspected clusters are below 10$^4$ M$_{\odot}$ (right column in the figure).
We see that the number of clusters in NGC~45 is small, 
compared with the other three galaxies. To first order, this appears to be consistent with the
overall low star formation rate derived for this galaxy.
Clusters in all four galaxies display a range in age and mass, but most are younger than 1 Gyr
and have masses below $10^5$ M$_\odot$, with few exceptions. We cannot exclude, however, that
the sample contains some older clusters that have been assigned too young ages. In particular,
relatively metal-poor old globulars would not be fit well by the models used here.
Our sample includes two of the eight spectroscopically confirmed old globular clusters in NGC~45 from
\citet{mora08}. From these we find ages of 0.04 and 0.8 Gyr, while Mora et al. find
spectroscopic average ages of 4.5 and 6.5 Gyr but consistent with ages as old as ~10 Gyr. This confirms 
the suspicion that some of the clusters in our sample might be older, metal-poor globulars with misassigned ages.

Our catalogs of the cluster candidates are available online. Table \ref{tab:olm} shows a few 
example lines to illustrate the format and the information contained in the catalogs.

\subsection{Cluster disruption and formation efficiencies}

Although we have derived masses and ages for our cluster candidates, some additional steps are
necessary before we can use this information to derive cluster formation rates. In Paper I we 
used a Schechter \citep{schechter76} mass function ($M_{\star}=2\times10^5$ M$_{\odot}$) to
extrapolate below the (age-dependent) mass limit and in this way we estimated the total mass
in clusters with ages between $10^7$ and $10^8$ years for NGC~4395. This approach ignores any
effects of disruption but is still useful for relative comparisons. We therefore first apply the
same approach to the four galaxies in this paper. Completeness limits were estimated by
plotting the luminosity functions and identifying the point where they start to deviate
significantly from a smooth power law. This occurs at the following absolute magnitudes: 
M$_{V}=-6.2$ for NGC~5236, M$_{V}=-5.7$ for NGC~7793, M$_{V}=-6.8$ for NGC~1313, and M$_{V}=-5.9$ 
for NGC~45. After estimating the mass in clusters with $10^7 < \tau/{\rm yr} < 10^8$ down to
a limit of 10 M$_{\odot}$ and dividing by the age interval (see Paper I for details), the resulting CFRs were
normalized to the area of the full ACS fields (a factor of $\sim2.27$ more) for comparison with 
the field star formation rates. The resulting CFRs are listed in the second column of 
Table \ref{tab:cluparam} (CFR$_{\rm P1}$).

Especially for NGC~45, the CFRs derived in this way are highly uncertain owing to the small number of clusters
that have four-band photometry. Better statistics can be obtained by only using the three-band
photometry in the ACS frames, but at the cost of having no age information for individual
clusters. However, CFRs may still be estimated by comparing the observed \emph{luminosity
functions} (LFs) with scaled model LFs \citep{gieles09}. If the CFR is assumed constant
and assumptions made about the initial cluster mass function ($\Psi$) and disruption parameters,
the LF can be modeled as follows \citep[Eq. 7 in][]{larsen09}:

\begin{figure}[!t]
	\centering
		\includegraphics[trim= 0mm 0mm 0mm 0mm,width=\columnwidth]{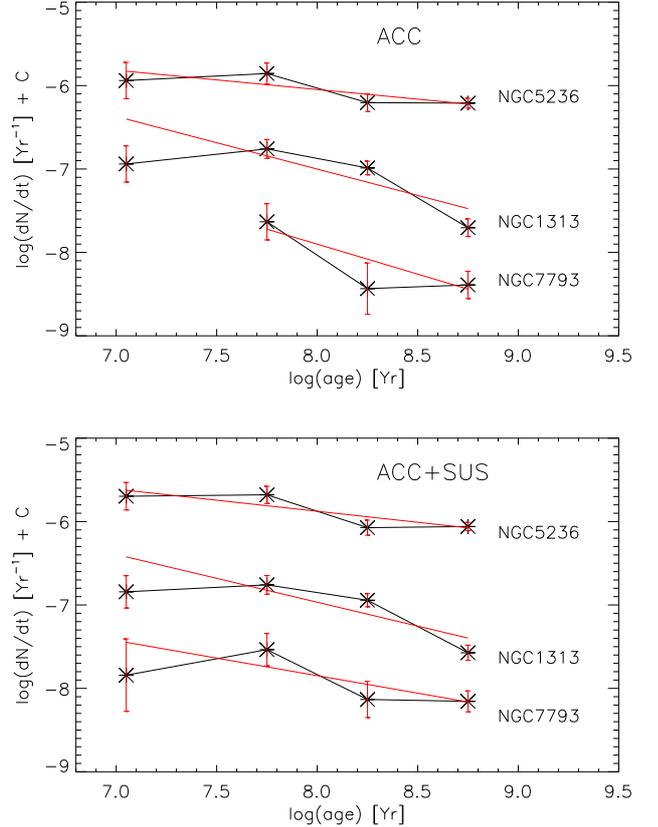}
		\caption{Age distributions for clusters in the galaxies NGC~5236, NGC~7793, and NGC~1313 in 
		              the mass range 10$^4$ to 10$^5$ M$_{\odot}$. Black lines represent the distributions for galaxies.
			    Red lines represent the best fit. Upper panel are the ADs only using the clusters classified as accepted.
			    The lower panel are the ADs for the clusters classified as accepted plus suspected.}
	\label{fig:agedist}
\end{figure}
\begin{eqnarray}
\frac{dN}{dL} & = & \int_{\tau_{min}}^{\tau_{max}} \Psi_i[M_i(L,\tau)] \times \frac{dM_i}{dM_c}\times \Upsilon_c(\tau) \times {\rm CFR} \nonumber \\
                        &    &  \ \ \ \ \ \ \ \ \ \ \ \ \  \times f_{\rm surv}(\tau)  \ d\tau \ ,
\label{eq:dndl}
\end{eqnarray}
where $\Psi_i[M_i(L,\tau)]$ is the initial mass function; $\Upsilon_c(\tau)$ is the mass-to-light ratio, which is only dependent on time in order to
be able to compute it from classical SSP models; CFR is assumed constant over time; and $f_{\rm surv}(\tau)$ is the number of clusters that 
survive after applying MID; $M_i$ and $M_c$ are the initial and current masses of a cluster
with luminosity $L$ and age $\tau$, and these are related through the assumed secular disruption law \citep{lamers05}.
The description of MID adopted here assumes that a constant fraction of the cluster population is removed
per logarithmic age bin, as opposed to a constant fraction of the mass of each individual cluster. If
$\Psi_i$ is a uniform power law, this makes no difference. However, if the MID stem from gradual mass loss
from individual clusters, any features in $\Psi_i$ (such as the
cut-off mass for the Schechter function) will be shifted downwards with time, whereas only the normalization
of $\Psi_i$ will change with time for constant number loss. 
% and $\frac{dM_i}{dM_c}$ is the relation between the cluster {\em initial} mass and the cluster {\em current} mass at time $\tau$ \citep[see Eq. (4) in][]{larsen09}, 
% which takes into account secular disruption.

In order to apply Eq.~(\ref{eq:dndl}), some constraints on cluster disruption are necessary. 
Models and empirical constraints on cluster disruption have been discussed in recent years by different 
authors \citep[e.g.][ among others]{BL03,lamers05,whitmorechandarfall07,larsen09,zhang99,fall04}, and \citet{fall09}
for different types of galaxies such as the LMC, SMC, Milky Way, M83, or Antennae. 
Given that it is currently uncertain to what extent MID or MDD dominates the cluster disruption, we 
carried out our analysis for both scenarios. 

\begin{table}[!t]\centering
\caption{Slopes for the age distributions for the accepted and the accepted+suspected samples.}
\begin{tabular}{c c c c}
\hline \hline
Mass [M$_{\odot}$]  & NGC~5236 & NGC~7793 & NGC~1313 \\ \hline 
\multicolumn{4}{c}{Accepted} \\
$10^4-10^5$  & -0.23$\pm$0.1 & -0.72$\pm$0.27 & -0.63$\pm$0.12 \\ \hline
\multicolumn{4}{c}{Accepted+Suspected} \\
$10^4-10^5$  & -0.26$\pm$0.08 & -0.42$\pm$0.19 & -0.57$\pm$0.11 \\ \hline
\label{tab:agedist}
\end{tabular}
\end{table}

Figure \ref{fig:agedist} shows the age distributions (ADs) for clusters with masses between $10^4$ and $10^5$ M$_\odot$ for
the galaxies NGC~1313, NGC~5236, and NGC~7793. NGC~45 has too few clusters to derive meaningful ADs.
We show fits for both the Accepted
and Accepted$+$Suspected samples. The slopes of the age distributions, obtained
by carrying out fits of the form $\log(dN/dt)=a\times \log(\tau)+b$ to the data
in Fig.~\ref{fig:agedist}, are given in Table \ref{tab:agedist}. There are no large differences
between the slopes derived for the Accepted and Accepted$+$Suspected sample. 
Figure \ref{fig:agemass} shows that most of the clusters in the suspected sample
have masses below our limit of $\log(M)$ [M$_{\odot}]=4.0$, explaining the similarity of
the age distributions above this limit. 

As a consistency check for the slope of the age distributions, we performed a maximum likelihood
fit to the data, assuming a power-law relation and using the power-law index as a free parameter. Using the accepted and accepted plus suspected sample of clusters,
we estimated the slope of the age distributions using the same age and mass ranges as shown in
Fig. \ref{fig:agedist} (i.e. ages between 4 Myrs up to 1 Gyr and masses between $10^4$ and $10^5$ M$_{\odot}$).
The results obtained are presented in Table \ref{tab:mls}. The derived slopes agree very well with those in table \ref{tab:agedist}, within the errors.
%We do not observe a significance variation
%over the new estimations of the slope of the age distributions. 

Using clusters with ages between $10^{6.6}\le \tau \le10^8$ yr and a mass $10^4\le M \le10^5$ M$_{\odot}$ we checked for (possible) variations over the slope of the age distributions.
The best fit for these slopes in the new age interval are $-0.17\pm0.39$, $-0.38\pm2.34$, and $0.78\pm0.43$ using the
accepted sample, and $-0.39\pm0.31$, $0.05\pm0.91$, and $0.59\pm0.91$ using the accepted plus suspected sample, for the 
galaxies NGC~5236, NGC~7793, and NGC~1313 respectively. The new slopes are to be flatter
than the values for the whole age range, showing even positive values; however, in most of the cases, the error is larger than the
estimation itself. The possibility of having shallower slopes indicate that there is a possible curvature of the age distribution
of star cluster systems, which is not consistent with a (simple) power law. 
%For clusters younger than 100 Myr the sample becomes deeply affected
%by the number of clusters to be used for the estimation of these slopes, especially for NGC~7793. 
%We emphasis that reliability of these results is questionable due to the low number of clusters used for these estimations and
%the fore-coming interpretations must be taken with caution.  

\begin{table}[!t]\centering
\caption{Slopes for the age distributions for the accepted and the accepted+suspected samples based on the maximum likelihood fit.}
\begin{tabular}{c c c c}
\hline \hline
Mass [M$_{\odot}$]  & NGC~5236 & NGC~7793 & NGC~1313 \\ \hline 
\multicolumn{4}{c}{Accepted} \\
$10^4-10^5$  & -0.27$\pm$0.11 & -0.48$\pm$0.32 & -0.63$\pm$0.10 \\ \hline
\multicolumn{4}{c}{Accepted+Suspected} \\
$10^4-10^5$  & -0.31$\pm$0.09 & -0.30$\pm$0.21 & -0.60$\pm$0.09 \\ \hline
\label{tab:mls}
\end{tabular}
\end{table}

The slopes found here are, however, somewhat shallower than the value of $a=-0.9\pm0.2$ found for NGC~5236
by \citet{chandarwhitmore10}. If interpreted within the MID scenario, the slopes derived here correspond 
to MID disruption rates of 41\%, 81\%, and 77\% per decade in age if we use the accepted sample, 
and 45\%, 62\%, and 73\% if we use the sample of accepted plus suspected objects, 
for the galaxies NGC~5236, NGC~7793, and NGC~1313, respectively. A weighted average of the slopes
for the Accepted samples leads to a mean slope of $\langle a \rangle = -0.42\pm0.07$ and
an MID disruption rate of $(62\pm6)$\% per decade in age. We use this mean weighted value %1-10^(-0.42) ± alog(10)*(1-10^(-0.42))*(0.01)
for all the galaxies in our sample, including NGC~45 and NGC~4395 where the numbers of detected
clusters are too low to allow us to estimate the slopes of the age distributions independently.

It is worth comparing our age distributions with those of \citet{mora09}. Their slopes were
considerably steeper than those found here, but this is partly because Mora et al.
worked with magnitude-limited samples, while our age distributions are for mass-limited samples.
When taking this into account, Mora et al. found that their data were consistent with
MID disruption rates of 75\%--85\% per dex, still somewhat higher than the values found here.
We note that the stochastic effects, combined with our relatively conservative size cuts, likely 
cause us to underestimate the number of objects in the youngest bins. This might account for some 
of the flattening of the age distributions in the youngest bins that is also seen in Fig.~\ref{fig:agedist} for 
NGC~1313 and NGC~5236. For this reason our estimates of the slopes and disruption rates might 
also be considered lower limits. The less strict size cut used by Mora et al. (FWHM=0.2 pixels
instead of our 0.7 pixels) would cause them to detect more compact, young
objects, but also possibly ones with increased contamination. Finally, since visual selection was
part of the sample selection in both our work and the one of Mora et al., this will also
lead to differences in the final samples.

\begin{figure}[!t]
	\centering
		\includegraphics[trim= 0mm 0mm 0mm 0mm,width=\columnwidth]{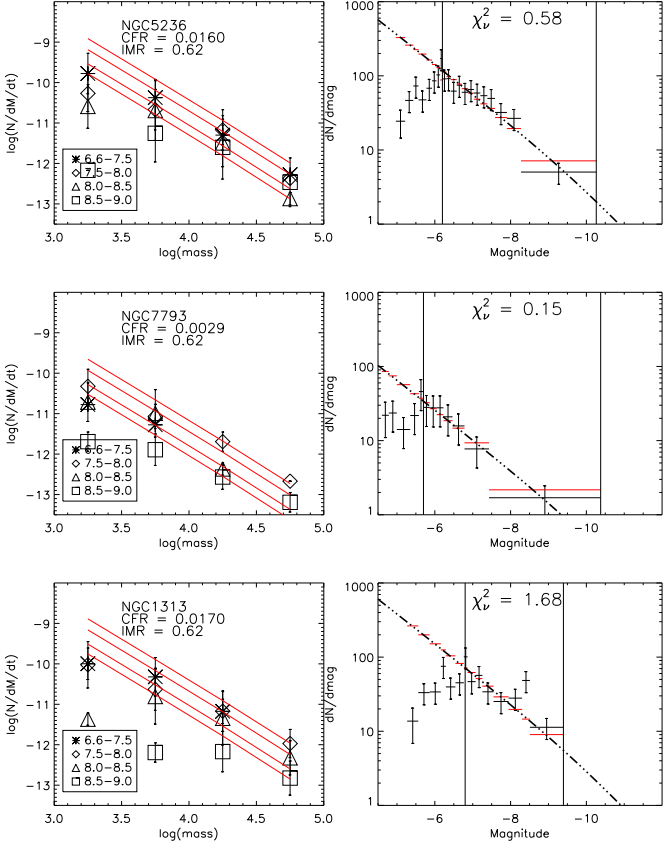}
		\caption{Mass-independent models. Left column: MDs observed (black symbols) and predicted models (red lines). 
%			      Center Column: LDs observed (black symbols) and predicted models (red lines). 
			      Right column: Luminosity function observed (black horizontal lines), predicted models (red horizontal lines), and theoretical model (dash-dotted line). 
			      Vertical straight lines represent the limits used for the fit of the LFs.}
	\label{fig:baltimore}
\end{figure}

\begin{figure}[!t]
	\centering
		\includegraphics[trim= 0mm 0mm 0mm 0mm,width=\columnwidth]{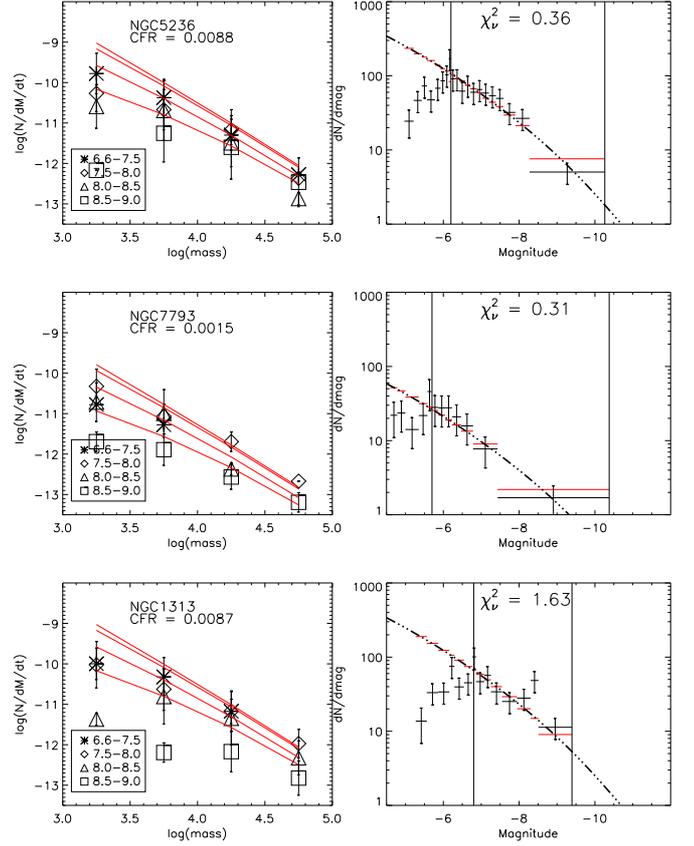}
		\caption{Mass-dependent models. Left column: MDs observed (black symbols) and predicted models (red lines). 
%			      Center Column: LDs observed (black symbols) and predicted models (red lines). 
			      Right column: Luminosity function observed (black horizontal lines), predicted models (red horizontal lines), and theoretical model (dash-dotted line). 
			      Vertical straight lines represent the limits used for the fit of the LFs.}
	\label{fig:utrecht}
\end{figure}

A key difference between the MID and MDD models is that, while the former predicts no change
in the shape of the mass distribution with age, the latter predicts a flattening at
low masses. In order to test whether we can distinguish between the two scenarios, model
mass distributions (MD, number of objects, per mass bin, per linear age bin) were computed for different
age intervals using a relation analogous to Eq. \ref{eq:dndl}, but integrating over $\tau$ for fixed $M$ rather 
than fixed $L$. These model MDs were then compared with the observed MDs in the same age intervals.
%We also computed luminosity distributions (LD, number of objects, per magnitude bin, per linear age 
%bin) for the same age intervals. For the latter we used $\Upsilon_c$ values from the GALEV SSP models and we
%assumed the mass functions to be of the \citet{schechter76} form with a low-mass slope of $\alpha=-2$. 
For the MDD scenario we assumed $\gamma=0.62$ and a disruption time of $t_4 = 1\times10^9$ yr \citep{lamers05}.
No infant mortality was included in the MDD models (i.e. $f_{\rm surv} = 1$ at all ages).
The MID models used the IMR obtained above (62\%), which is active in the age range 5 Myrs to 1 Gyr, 
and $t_4$ was set to infinity, making the models independent of mass.
We compared the model MDs with our observations, dividing our cluster samples into the 
following age and mass bins: log(M)[M$_{\odot}$]=[3.0,3.5,4.0,4.5,5.0] and log($\tau$)[yr]=[6.6,7.5,8.0,8.5,9.0] 
(shown in Fig. \ref{fig:agemass}).   

\begin{figure}[!t]
	\centering
		\includegraphics[trim= 0mm 0mm 0mm 0mm,width=\columnwidth]{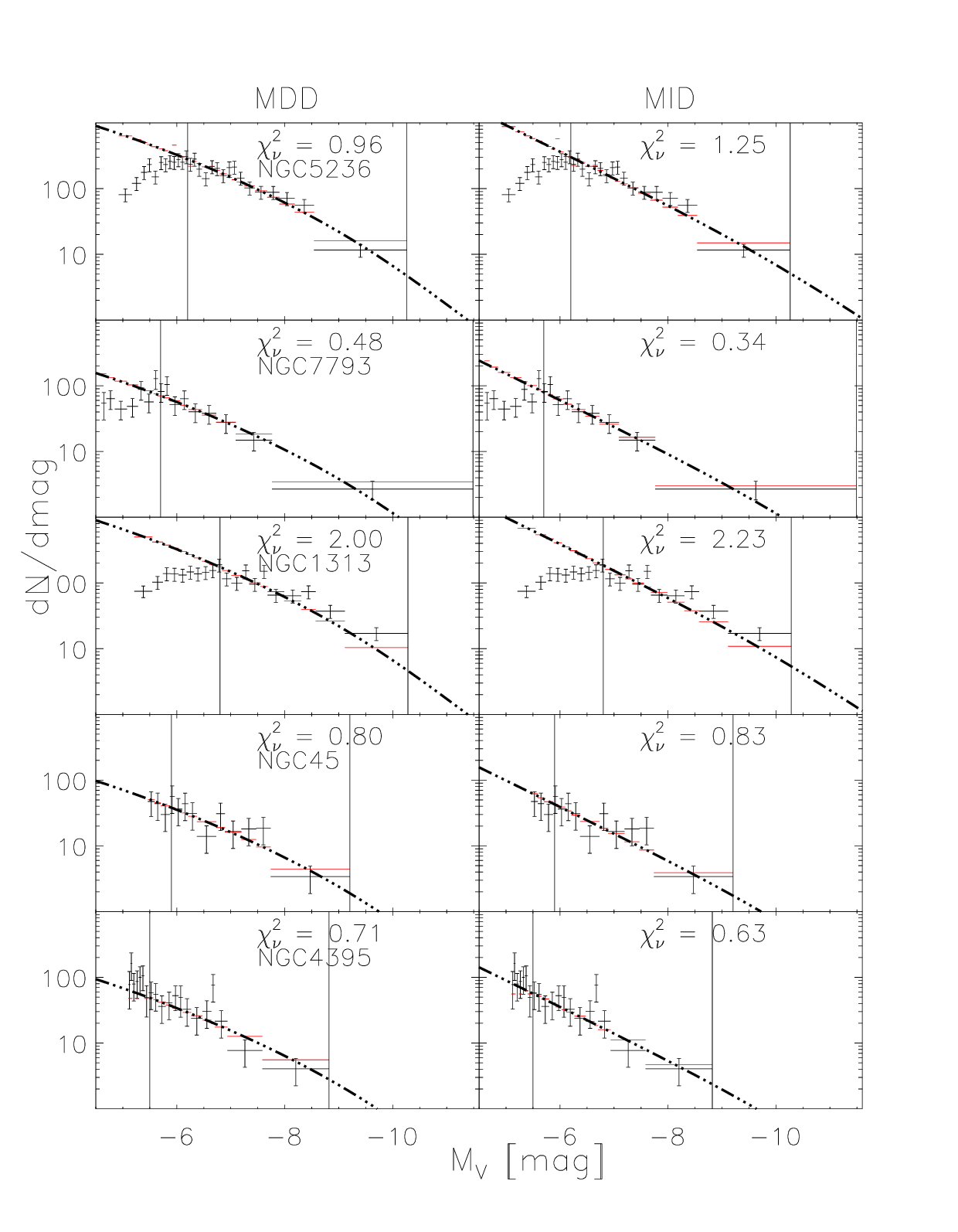}
		\caption{LFs for the five galaxies in our sample. We used MDD and MID theoretical models in left and right columns, respectively. Vertical lines are the limits of the fit.
			      Red horizontal lines represent the binned theoretical values, while black horizontal lines represent observations. Errors are Poissoninan. Dashed-dotted line
			      is the theoretical LF not binned.}
	\label{fig:lfs}
\end{figure}

In Figs. \ref{fig:baltimore} and \ref{fig:utrecht} we compare the observed and predicted MD and LF for 
MID and MDD models. The best fits were defined by scaling the model LFs to match the observed ones, where the 
observed LFs were created using variable binning with ten objects per bin for NGC~1313 and
NGC~5236 (following \citet{maizapellaniz09}) and five objects per bin for NGC~45, NGC~4395, and
NGC~7793. Errors were assumed to be Poissonian. Also shown is the reduced $\chi^2$ for each galaxy. 
Based on these figures, it is difficult to distinguish between the MID and MDD scenarios, mostly
because of the limited dynamical range of the data and the poor statistics. 
%However, MDD models
%with much shorter dissolution time scales than the assumed $t_4 = 1$ Gyr appear to be ruled out,
%as no clear flattening of the observed MFs is seen at low masses.

\begin{figure}[!t]
	\centering
		\includegraphics[width=\columnwidth]{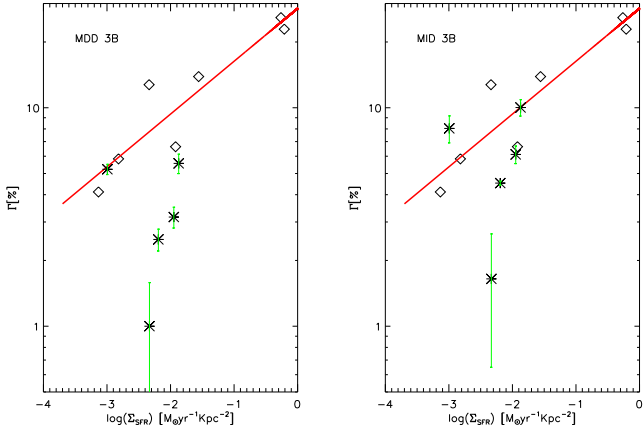}
		\caption{Relation between $\Gamma$ and $\Sigma_{SFR}$ using average values of $\Gamma$ for the models MDD and MID. Rhombs symbols and line represents the \citet{goddard10}
			      data and black star symbols represent our set of galaxies.}
	\label{fig:gammas}
\end{figure}

Our final estimates of the CFRs come from the LFs for clusters with 3-band photometry in
all five galaxies in our sample, using model LFs computed for both MDD and MID disruption.
The observed LFs were created in the same way as described above, using variable binning. 
We used the magnitude ranges up to the brightest cluster in the 
sample and the lower limit was set at the completeness limits estimated above (see vertical lines in Fig. 15).
The CFRs are listed in Table 7 for the 4- and 3 band photometry.
The CFRs for the four-band photometry have been scaled
to the size of the ACS fields. The observed LFs and the
best-fitting models are shown in Fig. \ref{fig:lfs}, where MID and
MDD models have been tested for the five galaxies in our sample.
The CFRs inferred from the fits in Fig 15 only include the accepted clusters. If we include the ``suspected" 
objects then the CFRs increase by 52\%, 62\%, 30\%, 4\%, and 25\% for NGC 5236, NGC 7793, NGC 1313, 
NGC 45 and NGC4395, respectively, for the MDD scenario. Similar changes occur for MID. Furthermore, if we 
exclude the known ancient GCs in NGC~45 \citep{mora08} from the sample, the $\Gamma$ value for this 
galaxy decreases by $\sim20$\%. Taken together, this then makes NGC~45 less of an outlier in Fig. \ref{fig:gammas} and in 
general shifts the data points upwards.

\begin{table*}[!ht]\centering
\caption{Estimates of the cluster formation rates. Subscript P1 refers to estimations made following Paper I. 
	      Disruption models are labeled as MDD and MID. Number of bands used for the estimation are labeled as 3B and 4B.}
\begin{tabular}{c c c c | c c}
\hline \hline
Galaxy  & CFR$_{P1}$ &  CFR$_{MDD}^{3B}$ & CFR$_{MDD}^{4B}$ &  CFR$_{MID}^{3B}$ & CFR$_{MID}^{4B}$   \\ 
               &  [M$_{\odot}$yr$^{-1}$] & [M$_{\odot}$yr$^{-1}$]  & [M$_{\odot}$yr$^{-1}$] &  [M$_{\odot}$yr$^{-1}$] &  [M$_{\odot}$yr$^{-1}$]  \\ \hline
NGC5236  & 37.7$\times10^{-3}$  & 23.0$\times10^{-3}$  & 19.9$\times10^{-3}$  & 41.0$\times10^{-3}$  & 36.6$\times10^{-3}$     \\
NGC7793  & 14.5$\times10^{-3}$  & 4.0$\times10^{-3}$    & 3.4$\times10^{-3}$     & 6.8$\times10^{-3}$  & 6.6$\times10^{-3}$    \\
NGC1313  & 60.7$\times10^{-3}$  & 23.0$\times10^{-3}$  & 19.7$\times10^{-3}$  & 44.0$\times10^{-3}$  & 38.6$\times10^{-3}$    \\ 
NGC45      & 8.6$\times10^{-3}$     & 2.5$\times10^{-3}$    & 2.7$\times10^{-3}$  &4.4$\times10^{-3}$  & 3.6$\times10^{-3}$   \\ 
NGC4395 & 4.5$\times10^{-3}$     & 2.4$\times10^{-3}$    & 1.0$\times10^{-3}$   & 4.0$\times10^{-3}$  & 1.6$\times10^{-3}$   \\ \hline %REMEMBER TO CORRECT FOR AREAS (4BANDS)
\label{tab:cluparam}
\end{tabular}
\end{table*}

\begin{table*}[!ht]\centering
\caption{Estimations of $\Gamma$ and comparison with \citet{goddard10} results.}
\begin{tabular}{c c c c c | c c c | c c | c}
\hline \hline
Galaxy  &  $\Gamma_{P1}$ &  $\Gamma_{MDD}^{3B}$ &  $\Gamma_{MDD}^{4B}$ & $\Gamma_{MDD}\pm\sigma_{MDD} $ & $\Gamma_{MID}^{3B}$ &  $\Gamma_{MID}^{4B}$ & $\Gamma_{MID}\pm\sigma_{MID} $  & $\Gamma^{\dagger}$ & $T_L(U)^{\dagger}$ & $L_{CL}/L_{FS}$  \\
               &  [\%] & [\%]  & [\%] & [\%] &  [\%] &  [\%] &  [\%]  & [\%] & & [\%] \\ \hline
NGC5236  & 9.8   & 5.9  & 5.2 & 5.6$\pm$0.6 &10.5  & 9.4 & 10.0$\pm$0.9  & 10.3  & 2.36$\pm$0.31 & 9.0 \\
NGC7793  & 9.8   & 2.6  & 2.3 & 2.5$\pm$0.3 & 4.5    & 4.5 &  4.5$\pm$0.1 & 8.6       & 1.15$\pm$0.32 & 3.4 \\
NGC1313  & 9.0   & 3.3  & 2.9 & 3.2$\pm$0.2 & 6.4     & 5.7  &  6.1$\pm$0.6 & 9.9      & 1.49$\pm$0.44 & 11.7 \\
NGC45      & 17.3 & 5.0  & 5.4 & 5.2$\pm$0.3 & 8.8    & 7.2  &  8.0$\pm$1.1 & 2.2      & 0.24$\pm$0.17 & 6.1 \\ 
NGC4395  & 2.6   & 1.4  & 0.6 & 1.0$\pm$0.6 & 2.3    & 0.9  &  1.6$\pm$0.1 & 8        & 0.07$\pm$0.05 &  1.7\\ \hline
\label{tab:gammas}
\end{tabular}
	\flushleft{$^{\dagger}$\citet{LR00}. Super and sub-indices for $\Gamma$ are the same as Table \ref{tab:cluparam}. 
		      Columns 5 and 8 are the mean and standard error for each model. 
		      Column 9 values using Goddard et al. with our
		      $\Sigma_{SFR}$. Column 11 is the fraction between the light from clusters 
		      and the light from field stars in the magnitude range 18 to 23 using $M_V$.}
\end{table*}

\begin{figure}[!t]
	\centering
		\includegraphics[width=\columnwidth]{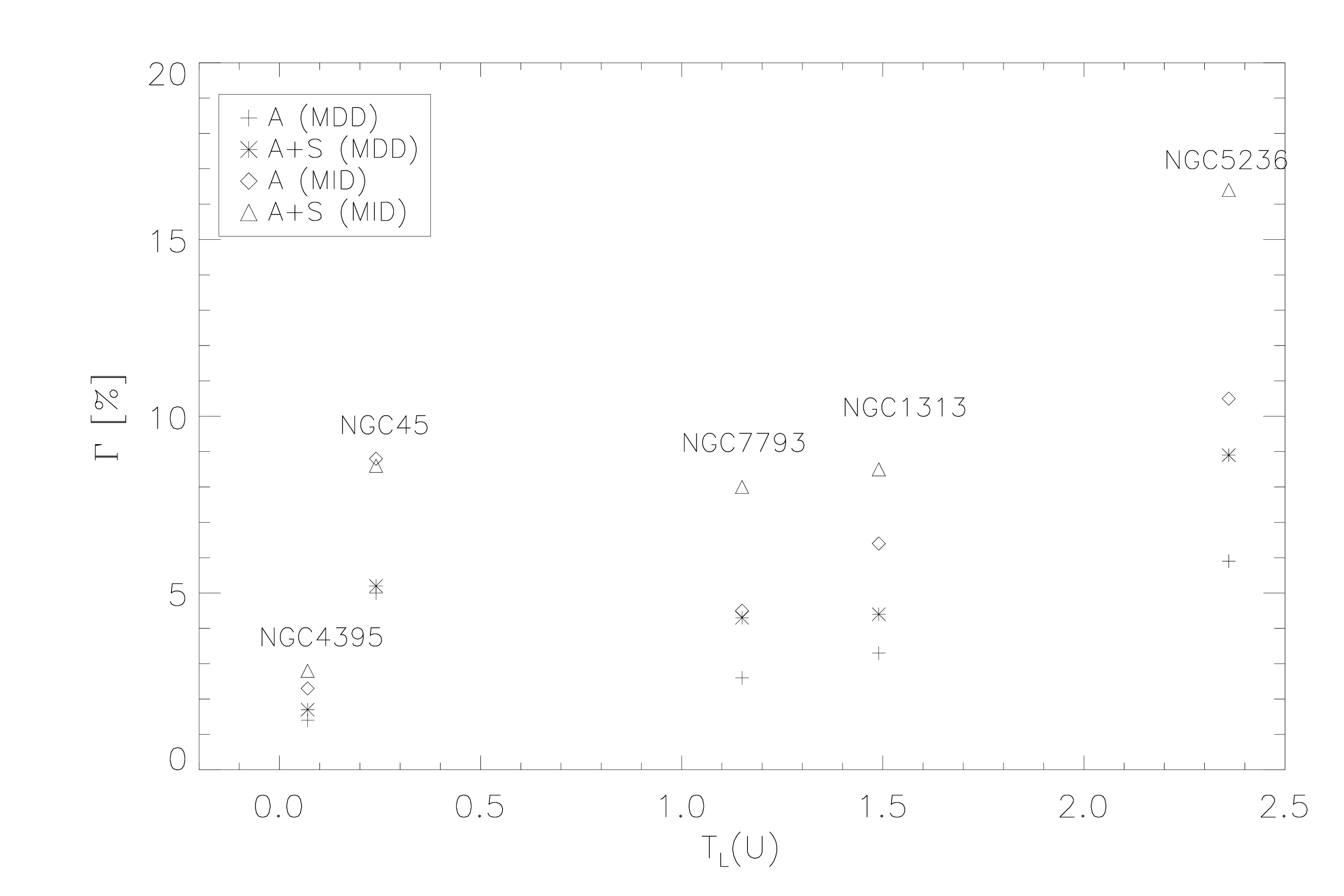}
		\caption{Relation between T$_L$(U) and $\Gamma$ using the models MDD and MID. The values for $\Gamma$ using the Accepted+Suspected sample are also included.}
	\label{fig:tlus}
\end{figure}

Having estimated the SFRs and CFRs, we are now able to calculate $\Gamma$. 
% SL: Removed the following, which is just a repetition of the introduction.
% \citet{gieles09} has claimed that this fraction depends 
% on environment, using the galaxies M74, M51, and M101. \citet{goddard10} have shown that there is 
% an apparent power-law relation between the SFR density ($\Sigma_{SFR}$) and $\Gamma$
% using 7 galaxies. In their work, the estimation of $\Gamma$ (and is relation with $\Sigma_{SFR}$) is based on the results of other 
% authors \citep[see Sect. 4 in][]{goddard10}, and the data sets do not cover the same mass and age ranges in all cases, although the authors corrected this issue
% extrapolating to a common lower mass limit. We decided to compare our results and use their data to see if the relation holds. 
Table \ref{tab:gammas} lists the $\Gamma$ values corresponding to the different estimates of the CFR. 
As an additional consistency check, we also include a direct comparison of the fraction of light 
coming from clusters and from field stars, using the magnitude range between m$_V=23$ and m$_V=18$. 
This is not a direct measure of $\Gamma$, but is still useful for checking any trends.
The $\Gamma$ values derived from the LF fitting are generally a few percent, with
the values derived from MDD models about 48\% smaller (or less) than those derived for the MID models.
We computed the mean value for each model and the respective standard error of the mean (columns 5th and 8th in Table \ref{tab:gammas}).
 
\section{Discussion}

In Fig. \ref{fig:gammas} we compare our $\Gamma$ vs. $\Sigma_{\rm SFR}$  measurements
with the data in \citet{goddard10} (rhombs). We cannot confirm whether a correlation is present
within the range of $\Sigma_{\rm SFR}$ values probed by our data, but overall our $\Gamma$ values are 
similar to those found by Goddard et al. in this range of $\Sigma_{\rm SFR}$ or slightly 
lower. However, following the methods of Goddard et al., \citet{adamo11} estimated values for
$\Gamma$ in the two blue compact galaxies ESO 185 and Haro 11. The results from 
Adamo et al. are in good agreement with the power law proposed by \citet{goddard10}.

According to \citet{LR00}, the five galaxies span a significant range in
specific $U$-band luminosity. From Table~\ref{tab:gammas} and Fig. \ref{fig:tlus}, we see that galaxies with
high $\Gamma$ values generally tend to also have high $T_L(U)$ values. One exception to this is NGC~45, which has a rather high $\Gamma$ for its 
$T_L(U)$. The $T_L(U)$ value for this galaxy is, however, based on only two
clusters, hence subject to very large uncertainty. Nevertheless, the
high $\Gamma$ value for NGC~45 is also somewhat puzzling given that it has the lowest
$\Sigma_{\rm SFR}$. This may suggest that there is not a simple relation between
$\Gamma$ and $\Sigma_{\rm SFR}$. In this context, it is also interesting that this
galaxy has a large number of ancient GCs for its luminosity, yielding an unusually
high globular cluster specific frequency for a late-type (Sd) galaxy \citep{mora09}. 

Our measurements of $\Gamma$ values in the range $\sim$2--10\% are consistent with
other recent estimates of the fraction of stars forming in bound clusters. It should be
kept in mind that this number is not necessarily an indicator of ``clustered''
vs. ``isolated'' star formation, since some stars may form in embedded clusters that
dissolve or expand on short enough time scales to drop out of our sample. 

There is a correlation between T$_{L}$(U) and $\Gamma$, as shown in Fig. \ref{fig:tlus}.
We estimated values for $\Gamma$ using the accepted+suspected samples for
the three-band photometry and for MDD and MID models as presented in the figure.
The galaxy NGC~45 deviates from the apparent relation. Two things must be
noted. (1) The inclusion of the suspected sample does not change the
trend observed dramatically, although the increases in the CFRs is reflected
in the new $\Gamma$ estimations, and (2) CFR estimates based on different disruption models follow the same
trend.

Estimates of the actual ``infant mortality rate'' are hard to make unless the embedded
phase is probed directly, something which is difficult in external galaxies. 
For NGC~1313, \citet{pellerin07} find that the IMR is a very efficient process 
for the dissolution of star clusters in this galaxy ($IMR=90\%$) based on UV fluxes 
in and out of clusters.  Our estimate of a $\Gamma$ value of 3\%--5\% for NGC~1313
indicates that $\ga95$\% of star formation in this galaxy happens outside clusters that 
are detected in our sample, in reasonable agreement with the Pellerin et al. estimate. However,
it is also clear that cluster dissolution is a continuous process, and systems that
are probed at older ages are generally expected to show a lower fraction of stars
in clusters. 

A proper account of dissolution effects could, in principle, be used to
correct measurements of $\Gamma$ at different ages to a common reference (say, 10 Myr), but
current uncertainties in the disruption process makes this difficult to apply in
practice. As a case in point, \citet{chandar10} estimate a mass-independent
disruption rate of 80\%--90\% per decade in age for NGC~5236,  a value 
that is significantly higher than our estimate $\sim40\%$ ($\sim62\%$ is our estimated weighted average). These
differences underscore that the definition of cluster samples (especially in
star-forming galaxies) is subject to strong selection effects, many of which
are age dependent and thus likely to affect the age distributions. One example
is the size cuts, which can easily cause a bias against young objects where
stochastic SIMF sampling leads to underestimated sizes even for masses
of $\sim10^4 M_\odot$. Furthermore, there may also be a physical relation between
cluster size and age \citep[e.g.][]{elson89,barmby09,mackeygilmore03,mackeygilmore2003}, compounding this problem.

When using young clusters ($\tau$ less than 100 Myr) we observed that the slope of the age distribution
gets shallower, indicating a possible curvature or deviation from a power-law relation and showing
values for the age distributions different from previous estimations. However, these results
are based on a sample that is strongly affected by very few of clusters.

%The differences in these values could be related with stochastic effects, which certainly changed the 
%number of objects detected, specially at young ages

%Following \citet{goddard10}, we decided to fit a line of the form $\Gamma=b\times\Sigma_{SFR}^{\alpha}$. Using only the 3 band estimations (for both models, MDD and MID)
%we estimated the slope for the two models used, and found $\alpha_{MDD}=0.26\pm0.09$ and $\alpha_{MID}=0.24\pm0.08$, and $b_{MDD}=22.36\pm0.04$ and $b_{MID}=22.87\pm0.18$, 
%in excellent agreement with the values estimated by Goddard et al. 

%Our results for $\Gamma$ can not constrain, neither answer, our question regarding if there is any relation between this quantity and the host galaxy. Values are affected by
%stochastic effects, which certainly changed the number of objects detected, specially at young ages, different models lead to different CFR (although values are similar). 
%We did not find a close relation between the SFR and $\Gamma$ based on our estimations, as suggested by \citet{goddard10}. 
%We estimated the mean value and the standard deviation for each model (see table \ref{tab:gammas}).

\section{ Summary and conclusions}
Using HST observations of the galaxies NGC~45, NGC~1313, NGC~4395, NGC~5236, and
NGC~7793, we studied their populations of star clusters and field stars separately
with the aim of constraining the quantity $\Gamma$, i.e.\ the ratio of stars
forming in bound clusters and the ``field''. We have been following the basic
approach described in Paper I, i.e. comparing synthetic and observed
color-magnitude diagrams (for the field stars) and SED model fitting (for the
star clusters), to get the formation histories of stars and clusters.

We tested how stochastic effects induced by the SIMF influence photometry and  
the estimation of ages and how the completeness limits are affected. We conclude that 
massive clusters (log(Mass)[M$_{\odot}$] $\ge 5$]) are easily detected (with the parameters used
in this work) at any age, while the detection of clusters with masses below
$\sim10^4 M_\odot$ can be strongly affected by stochasticity.
Our tests thus show that completeness functions do not just depend on magnitude, but
also on age and size. It would be desirable to find better classification methods
than a simple size cut to determine what is, and what is not, a cluster.

We estimated star formation histories  and found that
NGC~5236 and NGC~1313 have the highest star formation rates, while NGC~7793, NGC~4395,
and NGC~45 have lower SFRs. Within the uncertainties, we do not see significant
variations within the past 100 Myr.

Comparing observed and modeled mass- and and luminosity distributions for the cluster 
populations in different galaxies, we find that we cannot distinguish between
different disruption models (mass dependent vs.\ mass independent). We compared
model luminosity functions for each disruption scenario with observed LFs and
derived CFRs for the cluster systems. From our measurements of the CFRs and
SFRs we derived the ratio of the two, $\Gamma$, as an indication of the formation
efficiency of clusters that remain identifiable until at least $10^7$ years. We
find $\Gamma$ values in the range $\sim$2--10\%, with no clear correlation
with $\Sigma_{\rm SFR}$ within the (limited) range probed by our data. However,
our measurements are roughly consistent with those of \citet{goddard10}, who
find a relation between $\Sigma_{\rm SFR}$ and $\Gamma$ for a sample of galaxies
spanning a wider range in $\Sigma_{\rm SFR}$ (but more heterogeneous data).

A general difficulty in this type of work is to identify a reliable sample of
bona-fide clusters. Comparison with previous work suggests that the cluster samples, 
although covering the same galaxies, are significantly affected by the criteria used 
to classify clusters over the images. This results in different estimates of cluster system 
parameters, such as those related to the disruption law. Accurate estimates
of these parameters are also hampered by the relatively poor statistics that result
from having only patchy coverage of large, nearby galaxies in typical HST
imaging programs.

%%%%%%%%%%%%%%%%%%%%%%%%%%%%%%%%
%AKNOWLEDGEMENTS AND REFERENCES
%%%%%%%%%%%%%%%%%%%%%%%%%%%%%%%%

\begin{acknowledgements}
We would like to thank the referee for comments that helped to improve this article.
We would like to thank Morgan Fouesneau for catching an error in Fig. 8.
This work was supported by an NWO VIDI grant to SL.
\end{acknowledgements}

\bibliographystyle{aa}
\bibliography{../../../article_set}

\begin{landscape}

\begin{table}
\centering
\caption{Online material. The rows presented in this table illustrate what the online material will look like. {\em Column 1st}: Name of the galaxy, field observed and cluster number. 
	      {\em Column 2nd}: ID classification .{\em Columns 3rd and 4th}: X and Y coordinates of the clusters over the images. {\em Columns 5th and 6tõh}: Right ascension and 
	      declination (J2000). {\em Columns 7th-14th}: U, B, V, and I 
	      magnitudes and their respective errors. {\em Column 15th}: Age estimated for LMC-like metallicity. {\em Column 16th}: Mass estimated for LMC-like metallicity. {\em Column 17th and 18th}:
	      Sizes measured with SExtractor and Ishape, respectively. {\em Column 19th}: Flag for accepted and suspected objects.}

\begin{tabular}{c c c c c c c c c c c c c c c c }
\hline \hline
Galaxy\_Field\_\# & ID  & X & Y & RA & DEC & U & Ue & B & Be & V & Ve & I & Ie \\
              &  &  [PIX] & [PIX]  & [J2000] &  [J2000] & [Mag] &  [Mag] & [Mag] &  [Mag] & [Mag] &  [Mag] & [Mag] &  [Mag] \\ \hline
NGC5236\_1\_1 & 5214 &  245  &   2539. & 13:36:58.81& -29:51:13.94 & 21.08 &  0.10&  20.89  & 0.01 & 20.86  & 0.01 & 20.57  & 0.03 \\
NGC5236\_1\_2 & 5238 &  340  &   2549. & 13:36:58.99 &-29:51:09.77 & 22.90  & 0.21 & 22.27  & 0.02 & 21.66  & 0.02 & 20.64  & 0.02 \\
NGC5236\_1\_3 & 4360 &   596  &  2225. & 13:36:58.23 &-29:50:51.62 & 22.55 &  0.15 & 22.32  & 0.02 & 21.90  & 0.02  & 21.09  & 0.02 \\
NGC5236\_1\_4 &  5454 &  614  &  2609. & 13:36:59.62 &-29:50:58.36 & 20.73  & 0.04 & 20.64  & 0.01 & 20.54 &  0.01 & 20.14  & 0.01   \\
NGC5236\_1\_5 &  3919 &  621  &  1928. & 13:36:57.22 &-29:50:44.62 & 21.46 &  0.07 & 21.79 &  0.02 & 21.76  & 0.02 & 21.24  & 0.03\\

NGC5236\_2\_1 & 6116  &  574.  &  2690. & 13:37:05.64 &-29:56:51.97 & 22.07  & 0.08 & 22.15  & 0.01 & 22.04  & 0.01 & 21.61  & 0.02 \\
NGC5236\_2\_2 & 5845 &  655.  &  2423. & 13:37:04.84 &-29:56:42.65 & 22.05  & 0.13 & 21.65 &  0.02 & 21.40  & 0.01 & 20.71  & 0.02 \\
NGC5236\_2\_3 & 5690  &  719.  &  2180. & 13:37:04.10 &-29:56:34.62 & 22.29 &  0.12 & 22.57 &  0.02 & 22.28  & 0.02  & 21.62  & 0.02\\
NGC5236\_2\_4 & 5859  &  746.  &  2461. & 13:37:05.12 &-29:56:39.33 & 21.72  & 0.08 & 21.35  & 0.01 & 21.17  & 0.01 & 20.71  &  0.01 \\
NGC5236\_2\_5 & 5773  &  821.  &  2297.  & 13:37:04.67 &-29:56:32.47 & 21.87  & 0.08 & 21.97  & 0.01 & 21.74 &  0.01 & 21.23  & 0.02\\ \hline

\label{tab:olm}
\end{tabular}

\end{table}

% Continuation table
\begin{table}
\centering
\caption{Continuation Table \ref{tab:olm}.}

\begin{tabular}{c c c c c c }
\hline \hline
Galaxy\_Field\_\# & Log($\tau$) & Log(M) & FWHM$_{SEx}$ & FWHM$_{Isha}$ & Flag \\ \hline
 & [Yr] & [M$_{\odot}$] & [Pix] & [Pix] & \\ \hline
NGC5236\_1\_1 &  8.00  &  3.86 & 7.35 & 2.91 & acpt \\
NGC5236\_1\_2 &  8.87  & 4.43 & 5.84 & 0.74 & acpt \\
NGC5236\_1\_3 &  8.92  & 4.08 & 3.81 & 1.30 & acpt \\
NGC5236\_1\_4 &  8.39  & 4.20 & 4.80 & 1.93 & acpt \\
NGC5236\_1\_5 &  8.05  & 3.53 & 6.16 & 1.63 & acpt \\

NGC5236\_2\_1 &  8.00 &  3.48 & 3.89 & 0.95 & acpt \\
NGC5236\_2\_2 &  8.78  & 4.16 & 7.37 & 3.67 & acpt \\
NGC5236\_2\_3 &  7.81  & 3.48 & 3.91 & 1.35 & acpt \\
NGC5236\_2\_4 &  8.70  & 4.18 & 6.79 & 2.79 & acpt \\
NGC5236\_2\_5 &  8.39  & 3.73 & 4.39 & 1.84 & acpt \\ \hline
\label{tab:olmc}
\end{tabular}

\end{table}

\end{landscape}

\end{document}